\documentclass[prd,twocolumn,superscriptaddress,showpacs,floatfix,aps]{revtex4-2}
\pdfoutput=1
\usepackage[utf8]{inputenc}
\usepackage[english]{babel}
\usepackage[T1]{fontenc}
\usepackage{amsmath}

\usepackage{tikz}
\usetikzlibrary{tikzmark,calc,patterns}
\usepackage{makecell}

\usepackage{mathtools}
\usepackage{verbatim}
\usepackage{hyperref}
\usepackage{ulem}
\usepackage{tikz}
\usepackage{lipsum}
\usepackage{amssymb}
\usepackage{fixltx2e}
\usepackage{graphicx} 
\usepackage{physics}

\usepackage{xcolor}

\begin{document}
\title{Correlated Atom Loss as a Resource for Quantum Error Correction}

\author{Hugo Perrin}
\email{hugo.perrin@qperfect.io}
\affiliation{University of Strasbourg and CNRS, CESQ and ISIS (UMR 7006), aQCess, 67000 Strasbourg, France}
\affiliation{QPERFECT SAS, 67200, Strasbourg, France}

\author{Gatien Roger}
\affiliation{University of Strasbourg and CNRS, CESQ and ISIS (UMR 7006), aQCess, 67000 Strasbourg, France}
\affiliation{University of Luxembourg, Esch-sur-Alzette, Luxembourg}
\author{Guido Pupillo}
\affiliation{University of Strasbourg and CNRS, CESQ and ISIS (UMR 7006), aQCess, 67000 Strasbourg, France}
\affiliation{Institut Universitaire de France (IUF), 75000 Paris, France}
\email{pupillo@unistra.fr}

\begin{abstract}
Atom loss is a dominant error source in neutral-atom quantum processors, yet its correlated structure remains largely unexploited by existing quantum error correction decoders.
We analyze the performance of the surface code equipped with teleportation-based loss-detection units for neutral-atom quantum processors subject to circuit-level, partially correlated atom loss and depolarizing noise. We introduce and implement a decoding strategy that exploits loss correlations, effectively converting the \textit{delayed} erasure channels stemming from atom loss to erasure channels. The decoder constructs a loss graph and dynamically updates loss probabilities, a procedure that is highly parallelizable and compatible with real-time operation.
Compared to a decoder that assumes independent loss events, our approach achieves up to an order-of-magnitude reduction in logical error probability and increases the loss threshold from $3.2\%$ to $4\%$. Our approach extends to experimentally relevant regimes with partially correlated loss, demonstrating robust gains beyond the idealized fully correlated setting.
\end{abstract}
\maketitle
\section{Introduction}
Neutral-atom quantum processors have emerged as a leading architecture for large-scale fault-tolerant quantum computing~\cite{evered2023,bluvstein2024,finkelstein2024,saffman2010, browaeys2020, henriet2020, morgado2021, graham2022, anand2024, scholl2023, ma2023,cao2024, reichardt2025}. They provide long coherence times up to tens of seconds, flexible architectures with dynamically reconfigurable optical-tweezer arrays and scalability to tens or even hundreds of thousands of qubits. The realization of high-fidelity multi-qubit gates for neutral atoms mediated by Rydberg excitations \cite{jandura2022,pagano2022,evered2023,finkelstein2024} has recently opened the way to the realization of first breakthrough experiments in quantum error correction (QEC)~\cite{evered2023,bluvstein2024,finkelstein2024}, similarly to  superconducting qubits~\cite{acharya2025} and ions~\cite{egan2021, ryan-anderson2021, postler2022,dasilva2024}. 

As the field transitions from error-prone simulators to fault-tolerant quantum computers \cite{eisert2025}, attention has increasingly turned to designing QEC protocols that are tailored to the dominant noise mechanisms of neutral atom memories and processors. Alongside spontaneous emission from electronically excited Rydberg states, two types of error channels are characteristic of neutral atom platforms: (i) leakage out of the computational subspace and (ii) atom loss. Leakage  out of the computational subspace can be caused, for example, by spontaneous decay from
a highly-excited electronic Rydberg state to a
lower-energy state, population remaining in the
Rydberg state after a given quantum gate or
black-body radiation \cite{ma2023}, while atom loss can arise for example from 
heating that causes atoms to escape the optical
traps, collisions with background gas particles, or
anti-trapping during an excitation to a Rydberg
state~\cite{cong2022,bluvstein2022,evered2023,chow2024,kobayashi2026}. Both leakage and loss errors mostly occur during Rydberg-mediated multi-qubit gates, where a laser pulse excites two atoms for a few hundreds of nanoseconds \cite{evered2023, ma2023,graham2019}.

Although traditionally viewed as particularly harmful, recent works have shown that these processes can be at least partially harnessed. 
Rydberg leakage errors can be converted into atom losses using, e.g. the repulsive force of the trap. Loss events can in turn be identified using loss-detection units (LDUs)~\cite{gottesman1997,preskill1998,cong2022,suchara2015,chow2024,knill2005,stricker2020,moses2023}, which entangle each data qubit with an auxiliary atom to herald a loss or naturally detected by a direct measurement such as in the measurement-based quantum computing (MBQC) paradigms~\cite{baranes2026,yu2024, yu2025} or with a modified construction of the Steane syndrome extraction~\cite{baranes2026}. When detected, loss events can then be converted into {\it delayed} erasure errors -- i.e., errors where the faulty qubit is known, but the precise time of occurrence of the error is unknown within two consecutive LDU cycles -- which are substantially easier to decode than generic Pauli errors~\cite{suchara2015,wu2022,kubica2023,ma2023,omanakuttan2024,niroula2024,yu2024,chang2025,gu2024,gu2025}.
Scalability of fault-tolerant quantum computers will then benefit from a precise characterization of the underlying loss mechanisms and the development of advanced decoding strategies adapted to them.  

Most studies have so far assumed {\it independent} atom-loss models and demonstrated improved error-correction performance by employing specialized decoders tailored to {\it uncorrelated} atom loss~\cite{yu2024,ziad2024,baranes2026,perrin2025}.
However, the dominant source of loss and leakage for neutral atom stems from Rydberg-mediated gates that naturally induce \textit{correlated} loss events, 
where two atoms can be lost at the same time \cite{wu2022}. 
Despite their experimental relevance, such correlations have so far been overlooked in standard approaches to error correction.  An exception is the recent work~\cite{yu2025}, which has focused on correlated losses in the context of MBQC where the role of data and ancilla qubits are exchanged at each QEC cycle. In their framework, correlated loss events are particularly detrimental, degrading the distance code performance by a factor of 2 (from $d/2$ to $d/4$) when not properly treated. The authors show that with an appropriate decoder, the $d/2$ scaling typical of Pauli errors can be recovered.

\begin{figure*}
    \centering
    \includegraphics[width=0.88\linewidth]{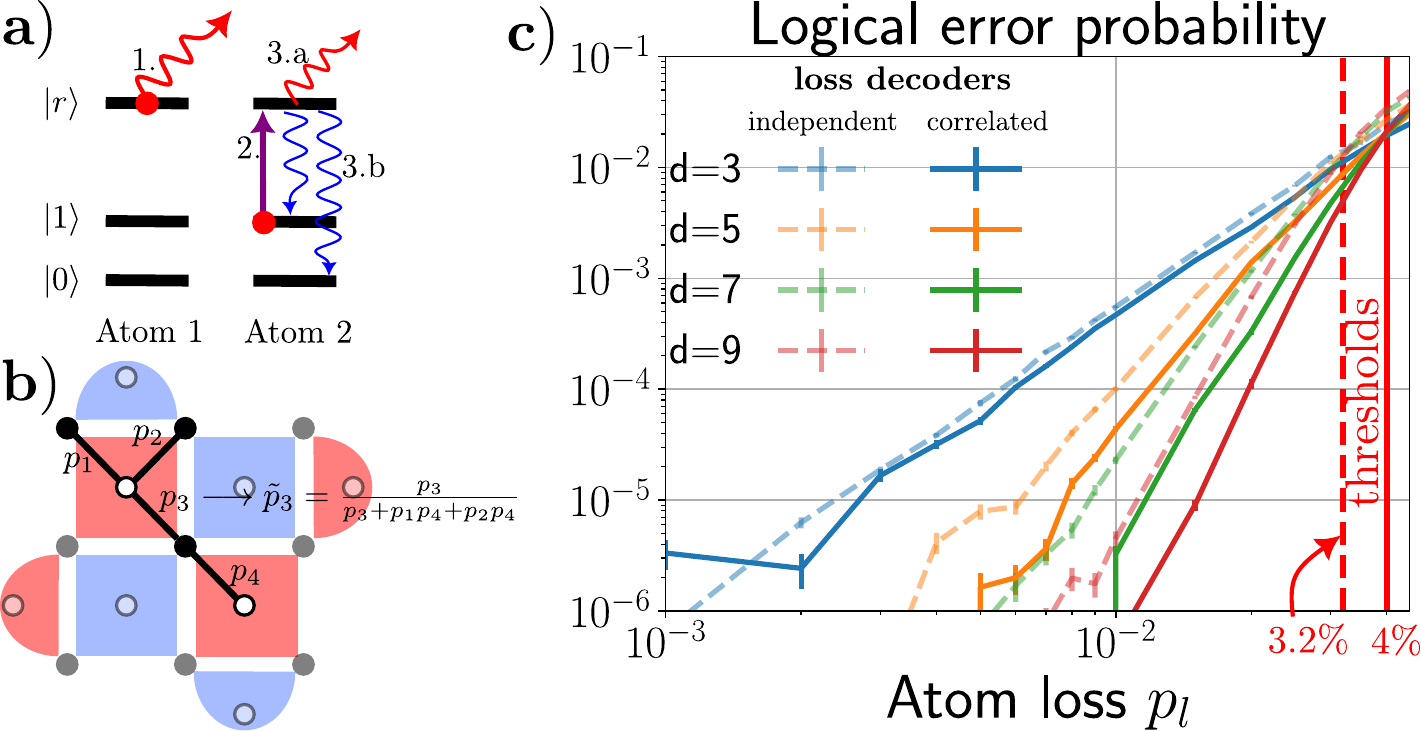}
    \caption{{\bf a)} Sketch of the correlated atom loss mechanism: an atom lost during the Rydberg pulse (step 1.) causes the remaining atom to be projected to $\ket{1}$ and potentially re-excited (step 2.), where it either undergoes subsequent loss (step 3.a) or decays back to the computational subspace (step 3.b). {\bf b)} Loss graph schematic: red/blue plaquettes denote $X/Z$ stabilizers; black/white dots are data/ancilla qubits; solid dots mark lost qubits (graph nodes). Edges connect simultaneously lost qubit pairs, with probabilities $p_i$ updated to $\tilde{p}_i$ by the fast correlated decoder based on the local neighborhood (see Eq.~\eqref{eq:update_proba}). {\bf c)} Logical error probabilities per round at vanishing depolarizing noise, $p_d = 0$ and for a fully correlated loss model $p_c=1$, as a function of the atom loss probability $p_l$ for code distances $d = 3, 5, 7, 9$ obtained by employing the independent-loss decoder (semi-transparent dotted lines) and the fast correlated loss decoder (solid lines). The dashed red vertical lines indicate the independent-loss decoder and the correlated loss decoder thresholds respectively at $p_l = 3.2\%$ and $p_l=4\%$.}
    \label{fig:figure_of_merit}
\end{figure*}

In this work, we show that the presence of correlations provides 
information that can be exploited to improve decoder accuracy and introduce a new decoder specifically designed to exploit these correlations. To benchmark our approach, we focus on the rotated surface code, however the approach can be extended to any other stabilizer-based codes, including quantum low-density-parity check codes \cite{pecorari2025a,bravyi2024}. The key idea of the decoder is to estimate the {\it a posteriori} probability of each loss mechanism conditioned on the loss syndrome. When these probabilities are strongly biased toward a particular loss mechanism, they effectively identify the most likely cause of the loss, thereby converting delayed erasure channels into standard erasure channels. To this end, we construct a loss graph, analogous to the Detector Error Model (DEM) used for Pauli noise, in which nodes are lost qubits and edges represent given loss mechanisms. We show that the {\it a posteriori} probability of loss mechanisms can be efficiently estimated using only local information from neighboring nodes in the loss graph (see Fig.~\ref{fig:figure_of_merit}{\bf b}).

For a fully correlated atom loss model, we demonstrate that the decoder improves the logical error rate up to one order of magnitude -- for a distance $d=9$ code -- compared to a decoder assuming independent atom loss, while also increasing the loss threshold from $3.2\%$ to $4\%$ (see Fig.~\ref{fig:figure_of_merit}{\bf c}). The scaling of the logical error is a power-law with exponent $d$ consistent with an erasure channel. 

The approach extends to experimentally relevant regimes with partially correlated loss, demonstrating robust gains beyond the idealized fully correlated setting. Interestingly, the proposed decoder is highly parallelizable and can operate in the sub-millisecond range making it compatible with real-time decoding in neutral atom quantum computers. 

This work has two main implications. First, correlated atom loss should not be regarded solely as a detrimental feature of the noise, but rather as a source of information that can be leveraged at the decoding stage. By exploiting these correlations, the decoder can infer the most likely underlying loss mechanisms and thereby effectively convert delayed erasure channels into standard erasure channels, restoring the advantage of having (probabilistically) identified error locations. Second, the results demonstrate the importance of incorporating realistic, circuit-level noise structure into decoder design. In particular, accounting for correlated loss processes through appropriate probabilistic inference—such as via the loss graph construction—yields substantial improvements in logical error rates and thresholds, underscoring the necessity of noise-tailored decoding strategies for scalable fault-tolerant quantum computation.

The article is organized as follows: in Section~\ref{sec:noise model}, we introduce the error model and describe the physical origin of correlated loss in neutral atom quantum computers. Section~\ref{sec:error correction} provides a brief overview of the rotated surface code and the LDUs. In Section~\ref{sec:decoder}, we review the standard decoding strategy for Pauli errors and independent qubit loss before presenting the correlated loss decoder. Finally, we benchmark its performance under circuit-level noise in Section~\ref{sec:logical error}.

\section{Noise model}
\label{sec:noise model}

In neutral-atom quantum computers, two-qubit gates are typically implemented by at least one laser driving the atoms to a given highly-excited electronic Rydberg level via a continuous pulse~\cite{jandura2022, pagano2022}. When atoms are sufficiently close to each other, the Rydberg blockade mechanism \cite{jaksch2000} prevents the atoms from being simultaneously excited to their respective Rydberg state due to strong van-der-Waals or dipole-dipole interactions~\cite{morgado2021, jandura2022, pagano2022}. The resulting interaction-induced energy shift generates entanglement, effectively realizing a CZ gate. 
This dynamics is described by the following Hamiltonian~\cite{jandura2022}
\begin{eqnarray}
    H&=& B\ket{rr}\bra{rr} \\
    &+&\Omega(t)\left(\ket{r}\bra{1}\otimes\mathbb{I}_2  +\mathbb{I}_1\otimes\ket{r}\bra{1}\right)+\text{h.c.}\nonumber
\end{eqnarray}
where $B$ is the interaction energy between the two atoms in the Rydberg state $\ket{r}$, $\Omega(t)$ is the (complex) time-dependent Rabi frequency of a laser incident on the atoms coupling states $\ket{1}$ and $\ket{r}$.  In the ideal case $B=\infty$, corresponding to a perfect Rydberg blockade, the state $\ket{11}$ partly evolves into the Rydberg-blockade state $\ket{W}=\frac{\ket{r1}+\ket{1r}}{2}$ while for the other states $\ket{00}$, $\ket{01}$, and $\ket{10}$, the qubit in state $\ket{0}$ remains unchanged and the qubit in state $\ket{1}$ is promoted to $\ket{r}$.

\subsection{Correlated loss mechanism}
During the two-qubit gate pulse, Rydberg-excited atoms are vulnerable to loss and leakage: Excited atoms are usually anti-trapped in the optical tweezer potential and can be lost during computation; they may also decay out of the computational subspace due to, e.g. spontaneous emission or coupling to nearby Rydberg states due to black-body radiation. We assume that population transferred outside the computational subspace either decays back to it between gate operations, or is permanently lost via anti-trapping or (auto-)ionization~\cite{wu2022}.

Here we consider that the loss of the first atom during the two-qubit Rydberg gate can increase the loss probability of the second atom, giving rise to {\it correlated-loss events} \cite{jandura2026,yu2025}: Assuming atom loss occurs exclusively when the qubit occupies the Rydberg state, the loss of the first atom projects the surviving atom into $\ket{1}$ for the component coming from the Rydberg–blockade state $\ket{W}$, or into $\ket{0}$ for the component originating from $\ket{10}$. The other components $\ket{00}$ and $\ket{01}$ are eliminated.

After the first atom loss and while the gate laser pulse is still being applied, there exists a finite probability that the atom in $\ket{1}$ is subsequently excited to its Rydberg state. As it undergoes only a partial evolution under the gate pulse, this second atom can remain in the Rydberg state at the end of the pulse and stay there until one of the following two processes occurs: (i) atom loss via leakage or anti-trapping, (ii) radiative decay back into the computational subspace (see Fig.~\ref{fig:figure_of_merit}{\bf a}).

The probability of losing the second atom is enhanced with respect to the first atom due to the longer effective occupation of the Rydberg state. The relative occurrence rates of processes (i) and (ii) above depend on the chosen atomic species and on the specific qubit encoding within the internal electronic levels of the atom. In particular, the branching ratios between decay into the computational subspace and decay channels leading to loss, as well as the branching between decay to $\ket{1}$ and $\ket{0}$, can vary significantly across experimental platforms~\cite{reichardt2025,bluvstein2022,bluvstein2024,tsai2025,radnaev2025,wu2022, ma2023}.

\subsection{Loss model}

We denote by $p_l$ the probability that an atom is lost during the implementation of a CZ gate. The atom subject to loss is chosen uniformly at random between the two atoms participating in the gate. If one of the two atoms had already been lost in  a previous operation before the gate starts, the remaining atom is considered lost with unit probability.
When both atoms are initially present, we introduce a second parameter, $p_c$, which quantifies the conditional probability of losing the second atom given that the first atom has already been lost during the same CZ gate. Under this model, the {\it marginal probability} for a given atom to be lost during the gate is
\begin{equation}
p_l^{\mathrm{marg.}} = \frac{p_l}{2}(1 + p_c).
\end{equation}
This expression follows from averaging over the two atoms, taking into account both the probability of being selected as the first lost atom and the additional contribution arising from correlated loss of the second atom.

If the second atom undergoes spontaneous decay back into the computational subspace, any phase coherence accumulated during its Rydberg excitation is lost, resulting in a phase-flip ($Z$) error. If the decay specifically populates the state $\ket{0}$, an additional bit-flip error is introduced.

Assuming an equal branching ratio of $50/50$ between decay to the two qubit states, we show in Appendix~\ref{ap:noise remaining} that, after Pauli twirling~\cite{kern2005,wallman2016,hashim2021}, the effective noise channel acting on the remaining atom can be described as a Pauli channel with error probabilities $1/8$ for both $X$ and $Y$ errors, and $3/8$ for $Z$ errors. Consequently, whenever the second atom is not lost, we model its state by applying the corresponding Pauli noise channel. We note that this noise model implicitly assumes that the remaining atom is always reexcited to the Rydberg state and subsequently undergoes either loss or a strong Pauli error. In practice, there is a finite probability that the atom is not reexcited and thus remains noise-free, which would weaken the effective Pauli noise channel. Our model therefore represents a worst-case scenario. \\

In Ref.~\cite{perrin2025}, Pauli noise channels acting on the remaining atom were analyzed in the regime of independent loss events, assuming a fully biased noise model in which errors occur exclusively as $Z$ errors. This situation corresponds to the limit where radiative transitions from the Rydberg state to $\ket{0}$ are forbidden. In such a regime, decoding performance is enhanced, as only a single type of Pauli error needs to be corrected.
In principle, this strong noise bias could be engineered experimentally by encoding the qubit subspace in appropriately selected electronic levels \cite{cong2022}. However, achieving this level of control remains experimentally challenging.

In this work, in the absence of atom loss, we model the noise following the CZ gate by applying a standard two-qubit depolarizing channel with error probability $p_d$. Single-qubit gates and measurements are assumed to be ideal, i.e., noiseless and unaffected by loss.

Finally, we emphasize that the loss mechanism is inherently state-dependent. In practice, loss events occur— or at least are significantly more likely to occur—when an atom is excited to the Rydberg state, and therefore primarily when it was initially prepared in the state $\ket{1}$ prior to Rydberg excitation. This state dependence is not explicitly incorporated into our simulations. In Appendix~\ref{ap:state dependent}, we further discuss this approximation in the context of quantum error correction. A more faithful implementation of a state-dependent loss channel, as well as potential adaptations of the decoding strategy to account for it, are left for future work.

\section{Error correction}
\label{sec:error correction}
In this section, we briefly review the error-correction strategy used to correct both standard Pauli errors and atom losses. Quantum error-correcting codes encode a logical qubit into many physical qubits in such a way that the logical information is delocalized and therefore resilient to local errors. In our work, we employ the widely used surface code, and more specifically its rotated variant, which requires roughly half as many qubits as the unrotated one.

To handle losses, the standard QEC cycle is augmented with LDUs that continuously reload fresh atoms into the processor. This mechanism keeps the total number of available qubits approximately constant throughout the computation, allowing the QEC protocol to operate reliably despite the presence of loss events.
\subsection{Rotated surface code}

Most quantum error-correcting codes are conveniently described in terms of their stabilizers $S_i$, which are operators that leave the encoded state $\ket{\psi}$ invariant in the absence of noise, i.e.,
\begin{equation}
\forall i,\,\,\,\, S_i\ket{\psi} =\ket{\psi}.    
\end{equation}
The stabilizer generators define a subspace — the codespace — in which the logical information is stored. For a code encoding a single logical qubit into $n$ physical (data) qubits, there are typically $n-1$ independent stabilizer generators, with the remaining degree of freedom corresponding to the logical qubit.

In practice, the stabilizers are repeatedly measured throughout the computation in order to detect and track errors. These measurements are implemented using additional ancilla qubits, which interact with the data qubits and are subsequently measured to extract the error syndromes without directly collapsing the logical state.\\

In the case of the surface code, the stabilizers consist of products of four Pauli $X$ or $Z$ operators (or two operators at the boundaries), placing it in the class of Calderbank–Shor–Steane codes. Its widespread adoption is largely due to its comparatively high error threshold (around $1\%$ for circuit-level noise) and low logical error rates, together with the fact that all required stabilizer measurements involve only local interactions on a two-dimensional nearest-neighbor lattice.

A key parameter of any QEC code is its distance $d$, defined as the minimum number of single-qubit errors required to implement a non-trivial logical operation and hence cause a logical error. For the rotated surface code, the distance scales as $d = \sqrt{n}$, where $n$ denotes the number of data qubits. For further details regarding the implementation and operation of the surface code, we refer the reader to the review~\cite{fowler2012}.

\subsection{Loss detection units}
In contrast to standard Pauli errors, atom losses cannot be corrected solely within the framework of conventional QEC codes. Intuitively, in the absence of a mechanism to replenish missing atoms, the computation would progressively involve fewer physical qubits, ultimately rendering error correction impossible. More formally, loss constitutes a non-Markovian noise process that is non-local in time, thereby violating the locality assumptions typically underlying fault-tolerant error-correction schemes.

To mitigate this issue, several strategies have been proposed. These include hardware-specific approaches~\cite{scholl2023,omanakuttan2024}, which physically replace lost atoms within the array, as well as circuit-based methods~\cite{cong2022,chow2024}, which restore the qubit while preserving the encoded logical information whenever the original atom has not been irreversibly lost.\\

In this work, we focus on an approach based on the teleportation-based LDUs, which has been shown to provide the best logical performance among existing proposals~\cite{chow2024,perrin2025,baranes2026}. In this scheme, each data qubit is replaced by a freshly prepared atom after each LDU. The quantum state of the original atom is transferred to the new one via a teleportation protocol requiring only a single entangling gate.

During the protocol, the original data qubit is measured. We assume that the measurement apparatus can distinguish between loss or leakage events and the computational states $\ket{0}$ and $\ket{1}$~\cite{senoo2025}. The measurement therefore yields three possible outcomes:

(i) If the data atom is present and the measurement outcome is $\ket{0}$ (which occurs with probability $50\%$), the quantum state is successfully teleported to the fresh atom without additional error.

(ii) If the measurement outcome is $\ket{1}$, the state of the fresh atom acquires a phase-flip ($Z$) error. Rather than correcting this error immediately via feedforward, it is accounted for at the decoding stage by appropriately flipping the outcomes of the neighboring $X$-type stabilizers.

(iii) If the original data atom has been lost, the LDU inserts a fresh atom initialized in state $\ket{0}$.\\

We note that replacing all data qubits during each LDU cycle also serves to cool the atomic array, thereby mitigating non-Markovian loss and leakage events that arise from heating. While this cooling effect is not explicitly included in our simulations, it represents an additional practical advantage of the teleportation-based LDU protocol~\cite{chow2024}.

\section{Decoders for correlated losses}
\label{sec:decoder}

In this section, we introduce a decoder designed to exploit correlations in atom-loss events. We begin by reviewing the standard decoding procedure for Pauli errors. Next, we describe the strategy used to handle independent atom losses. Finally, we present the principle underlying the correlated-loss decoder: we first introduce an accurate — but non-scalable — version, and then propose an approximate variant that significantly reduces the decoding runtime, is scalable and compatible with real-time operation in a quantum computer.

\subsection{Pauli noise decoder}
\label{sec:Pauli}

Quantum error correction relies on the repeated measurement of a code's stabilizers. When an error occurs, certain stabilizer outcomes flip, signaling the presence of an error. The decoder’s task is to infer the most likely set of errors that could have produced the observed syndrome — that is, the set of flipped stabilizers.

This decoding problem can be reformulated as a {\it perfect-embedded matching problem on a hypergraph}~\cite{higgott2025}. In this framework, the hypergraph is constructed as a {\it detector error model} (DEM): each possible error mechanism corresponds to a hyperedge, and activating a given error produces a set of flipped stabilizers, which correspond to the hypergraph's nodes. The probability of each error is encoded as the hyperedge weight, typically using the log-likelihood ratio of the error probabilities.

A "perfect matching" consists of selecting a subset of hyperedges that covers all nodes exactly once, while the term "embedded" refers to the situation where only a subset of nodes — corresponding to the observed syndrome — needs to be matched.

Among various decoding approaches, the minimal-weight perfect matching (MWPM) algorithm is widely used in QEC, particularly for the surface code, as it provides a favorable trade-off between decoding accuracy and computational latency. To apply MWPM, the DEM is first simplified into a graph-like structure—a transformation that is straightforward for the surface code. Software tools such as \texttt{Stim}~\cite{gidney2021} can automatically generate a DEM from a noisy Clifford circuit, and, together with the observed syndrome, the resulting matching problem can be efficiently solved using \texttt{Pymatching}~\cite{higgott2021,higgott2025}.

\subsection{Decoder for independent atom loss}
\label{sec:independent}
Reference~\cite{perrin2025} demonstrates how the decoding strategy discussed above can be extended to handle atom-loss errors in QEC. Analogous to Pauli errors, atom loss — and its subsequent replacement with a fresh atom via the LDU — alters the outcomes of neighboring stabilizers. Unlike Pauli errors, however, these stabilizer flips occur {\it probabilistically}, and multiple independent sets of stabilizers may be affected by a single loss event. As a result, each loss mechanism gives rise to its own DEM.

We note that the LDU provides additional information by signaling whether an atom has been lost. This process is analogous to an erasure channel, with the key distinction that the precise location of the loss between two consecutive LDUs is unknown~\cite{perrin2025} — a situation referred to as {\it delayed erasure} in Ref.~\cite{baranes2026}.

To leverage this information, the DEM associated with standard Pauli errors is augmented by incorporating the DEMs corresponding to atom-loss events. Constructing the DEM for a lost atom involves identifying all potential loss locations, generating the DEM for each, and weighting them by the probability that the loss occurred at a given location, conditioned on the detection of a loss.

A key underlying assumption in this procedure is that the DEM for multiple loss events can be obtained by a straightforward concatenation of the individual single-loss DEMs. While this assumption is exact for standard Pauli errors, it may only represent an approximation in the case of loss processes~\cite{liu2026}. Nevertheless, we note that Ref.~\cite{baranes2026} has recently shown that explicitly including two-body loss mechanisms in the DEM does not yield any improvement in decoding performance. Based on this observation, we adopt the same approximation in the implementation of the correlated-loss decoder described below.

\subsection{Accurate decoder for correlated atom loss}
\label{sec:accurate}

As discussed in Sec.~\ref{sec:noise model}, atom losses can be correlated when they occur during two-qubit Rydberg-mediated gates. These correlations carry information about the likely locations of losses, which can be exploited to improve the decoder’s accuracy. Here below, we first consider the case of fully correlated losses (i.e., correlation probability $p_c = 1$ in Sec.~\ref{sec:noise model}) and later extend the discussion to partially correlated losses.\\

As in the independent-loss model, the inclusion of a loss channel in the final DEM is conditioned on the absence of the corresponding atom, as indicated by the LDU or stabilizer measurements. However, unlike the independent case — where loss probabilities can be precomputed — these probabilities now depend {\it dynamically} on the entire current loss syndrome.

To illustrate this point and emphasize the importance of a correlated decoder, we present in Fig.~\ref{fig:loss_graph}{\bf a} a simple example where a delayed erasure can effectively be turned into a standard erasure channel:
\begin{figure}[!!h]
    \centering
    \includegraphics[width=\linewidth]{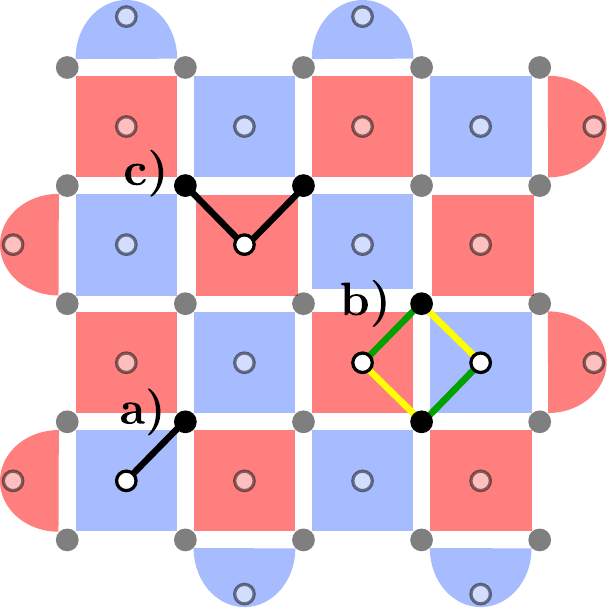}
    \caption{A schematic representation of the loss graph. Red (resp. blue) semi-transparent plaquettes correspond to $X$ (resp. $Z$) stabilizers, which detect $Z$ (resp. $X$) errors. Black (resp. white) dots represent data (resp. ancilla) qubits. Semi-transparent dots indicate qubits that remain present during the error-correction cycle, while solid dots denote lost qubits and define the nodes of the loss graph. An edge is drawn between two lost qubits whenever a physical error mechanism that lost simultaneously both qubits exist. {\bf a)}: A connected component of the loss graph consisting of two lost qubits {\bf b)}: A connected component consisting of four lost qubits, for which two distinct pairing configurations are possible, shown in green and yellow. {\bf c)}: A connected component consisting of three lost qubits. In this case, the only possible matching requires the ancilla qubit to be matched twice.
    The figure illustrates a single time slice corresponding to one cycle of the quantum memory experiment. To fully capture all loss mechanisms, the model must be extended to the time domain, since losses induced by the LDU CZ gates can connect qubits across consecutive error-correction cycles. }
    \label{fig:loss_graph}
\end{figure}
Consider two atoms lost during a single CZ gate within an error-correction cycle. Treating these losses as independent requires identifying all possible loss locations for both atoms and constructing a DEM that accounts for every loss scenario. However, if the losses are fully correlated—i.e., atoms can only be lost in pairs—the full loss syndrome immediately reveals that both atoms were lost simultaneously, precisely identifying the underlying loss location. Designing a decoder capable of exploiting such correlations is therefore essential.\\

In the simple case of Fig.~\ref{fig:loss_graph}{\bf a}, the loss location is identified unambiguously with probability one. For more complex syndromes, instead, multiple pairing configurations may reproduce the same loss pattern, as illustrated in Fig.~\ref{fig:loss_graph}{\bf b}, where two valid pairings (yellow and green edges) are consistent with the observed losses. To capture these correlations systematically, we construct a loss graph, with nodes representing lost atoms and edges connecting atoms that could be lost together via a CZ gate (see Fig.~\ref{fig:loss_graph}). Within this framework, determining the probability of a given loss mechanism reduces to a matching problem on the graph, as explained below.

The probability that a specific edge $e$ contributes to the observed loss pattern is given by
\begin{equation}
\tilde{p}_\text{e} = \frac{\sum_{s\in S_e} p_s}{\sum_{s\in S} p_s},
\label{eq:proba_loss}
\end{equation}
where $S$ is the set of all matching solutions for the syndrome, $S_e \subset S$ contains the solutions including edge $e$, and $p_s = \prod_{e' \in s} p_{e'}$ is the probability of a given matching, computed as the product of its edges’ {\it a priori} probabilities (for this model, $p_e = p_l p_c$). The resulting $\tilde{p}_e$ thus represents the {\it a posteriori} probability of each edge after matching.

In addition, even in the case of a fully correlated loss channel, it may not always be possible to pair all lost atoms. For instance, in Fig.~\ref{fig:loss_graph}{\bf c}, an odd number of atoms may be lost, implying that at least one atom cannot be paired. To address this issue, we allow nodes (i.e., lost atoms) to be matched multiple times. This situation corresponds physically to the case where an atom is lost during a two-qubit gate, while the other atom had already been lost in a previous operation.

Let $s$ be a solution to this generalized matching problem, and let 
$m_s(n)$ denote the number of edges in the solution that are incident to node $n$. We then define the {\it multiplicity} $k$ as
\begin{equation}
    k=\sum_{n\in V}(m_s(n)-1),
\end{equation}
which counts the number of times nodes are matched more than once—i.e., the total “excess” matching degree beyond a simple pairing. Each node matched exactly once does not contribute to the multiplicity $k$. We refer to such a configuration $s$ as a $k$-matching solution.\\

To accelerate the $k$-matching problem, we exploit the decomposition of the loss graph into {\it connected components}, which can be solved independently. For each component, only even (odd) $k$-matching solutions are possible if the number of nodes is even (odd). 

We begin by exploring small $k$ ($k=0$ for even, $k=1$ for odd), since larger $k$ increases the number of edges in the solution, reducing its overall probability. If no valid solution exists for a given $k$, we increment $k$ by two and repeat until a solution is found. When multiple $k$-matching solutions exist, the probability of each edge is computed using Eq.~\eqref{eq:proba_loss}. For a unique solution, the probability is one for edges in the solution and zero otherwise. The Pauli DEM is then augmented with the DEMs of each loss mechanism, weighted by these probabilities, and the resulting combined DEM is solved using \texttt{Pymatching}.

Rarely, no matching solution may reproduce the observed stabilizer syndrome - particularly if Pauli errors are absent. In this case, the $k$-matching search is repeated with $k$ incremented by two until a consistent solution is found.\\

The MWPM algorithm cannot be directly applied to the $k$-matching problem for two reasons: it returns only the minimal-weight solution and enforces perfect matchings. While graph modifications could allow MWPM to output additional or non-perfect solutions, here we employ a brute-force approach. Our implementation recursively explores all edges of each connected component, allowing nodes to be matched up to $k$ times. This ensures all $k$-matching solutions are found but scales exponentially with the number of edges, making it near-optimal in accuracy but impractical for large-scale or real-time decoding.

Partial correlations ($p_c<1$) are incorporated by adding edges connecting nodes to the vacuum, representing independent losses, while edges between nodes represent correlated losses. In our model — where LDUs are applied every cycle and correlated losses occur only during CZ gates — each edge between two nodes uniquely identifies a correlated loss channel. In contrast, vacuum edges may correspond to multiple independent-loss mechanisms, and their DEMs are constructed by weighting contributions according to their probabilities, as in~\cite{perrin2025}.

\subsection{A fast correlated loss decoder}
\label{sec:fast}
We now introduce a fast decoder that sacrifices a small amount of accuracy for a substantial gain in speed. Rather than computing exact loss-mechanism probabilities via the full $k$-matching algorithm, this approach estimates probabilities using only the local connectivity of the loss graph around each edge.\\
\begin{figure}[t!]
    \centering
    \includegraphics[width=\linewidth]{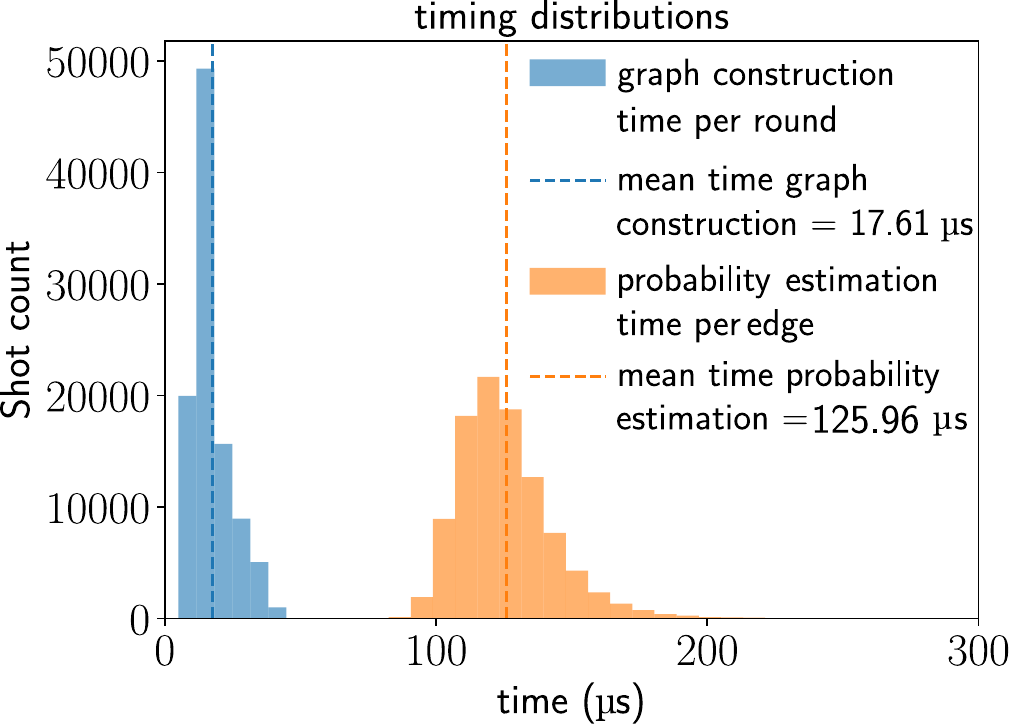}
    \caption{Distribution of the loss graph construction time normalized by the number of rounds $d = 9$ and the {\it a posteriori} probability estimation time normalized by the number of edges in the loss graph. Dashed lines indicate the mean of each distribution. These timing have been estimated for a surface code of distance $d=9$ with a partially correlated loss model ($p_l=0.01$, $p_c = 0.5$) at vanishing depolarizing noise $p_d=0$ over $10^5$ shots.}
    \label{fig:time}
\end{figure}

To estimate the probability that an edge $e$ — representing a specific loss mechanism — belongs to a valid matching, we consider the two nodes it connects, $n_1$ and $n_2$. These nodes can either be matched directly via $e$, or independently via other incident edges. The resulting approximate {\it a posteriori} probability is
\begin{equation}
\tilde{p}_e \simeq \frac{p_e}{\left(\sum_{\substack{e'\neq e,\\ n_1\in e'}} p_{e'}\right) \left(\sum_{\substack{e''\neq e,\\ n_2\in e''}} p_{e''}\right) + p_e}.
\label{eq:update_proba}
\end{equation}
This computation depends only on local edge weights and is fully parallelizable, as each $\tilde{p}_e$ can be evaluated independently. The DEM is then updated using these approximate probabilities, and the perfect embedded matching problem is solved in the same way as for the accurate decoder.

Despite its simplicity, the fast decoder achieves near-identical accuracy to the full $k$-matching decoder (see Fig.~\ref{fig:correlated_loss_no_depo} and discussion below), while operating in a way compatible with real-time decoding for neutral-atom quantum computers, under sufficient parallelization. In Fig.~\ref{fig:time}, we show the distribution of timings for the two steps that the fast decoder adds on top of the standard MWPM to handle correlated loss: the construction of the loss graph and the estimation of the {\it a posteriori} probabilities. These are obtained for a surface code of distance $d=9$ under a realistic loss model ($p_l=0.01$, $p_c=0.5$) with vanishing depolarizing noise $p_d=0$ (note that these timings are independent of the depolarizing noise strength). The loss graph construction time is normalized by the number of rounds  $d=9$, while the {\it a posteriori} probability estimation time is normalized by the number of edges in the loss graph, as these estimations are parallelizable. The latter also accounts for the time required to load the various loss DEMs, reweight them by their estimated probabilities, and combine them into a single file. In this regime, loss graph construction is highly efficient: timings are predominantly below  $50 \mu$s, with a mean of  $18 \mu$s. The {\it a posteriori} probability estimation time averages $126\mu$s, with most timings remaining under $200\mu$s. Together, these results confirm that the fast decoder comfortably satisfies the sub-millisecond-per-round latency requirement for real-time neutral-atom decoding.

\section{Logical error probability}
\label{sec:logical error}
In this section, we evaluate the performance of the rotated surface code supplemented with teleportation-based LDUs under depolarizing noise and (partial) correlated atom losses during CZ gates (see Sec.~\ref{sec:noise model}). Code distances $d = 3, 5, 7, 9$ are considered, with $d$ consecutive rounds of error correction per simulation.\\

For various combinations of depolarizing error probability $p_d$, loss probability $p_l$, and correlation probability $p_c$, we report the logical error probability per round~\cite{acharya2023,xu2024,pecorari2025b,perrin2025}, defined as
\begin{equation}
\varepsilon_r = 1 - (1 - \varepsilon)^{1/n_r},
\end{equation}
where $n_r = d$ is the number of rounds and $\varepsilon$ is the total logical error probability, obtained as the fraction of shots in which the decoder produces an incorrect logical correction. All simulations initialize the code in the logical state $\ket{0}_L$ and measure in the corresponding logical $Z$ basis.

A key figure of merit is the threshold, i.e., the physical error rate below which increasing the code distance yields exponential suppression of the logical error rate. Thresholds are indicated in the plots by a solid line.

Simulations are performed using Stim, with losses sampled gate by gate via Monte Carlo according to the model described in Sec.~\ref{sec:noise model}. LDUs are implemented following the procedure described in Ref.~\cite{perrin2025}.

\subsection{Fully correlated loss}

\begin{figure}[t!]
    \centering
    \includegraphics[width=\linewidth]{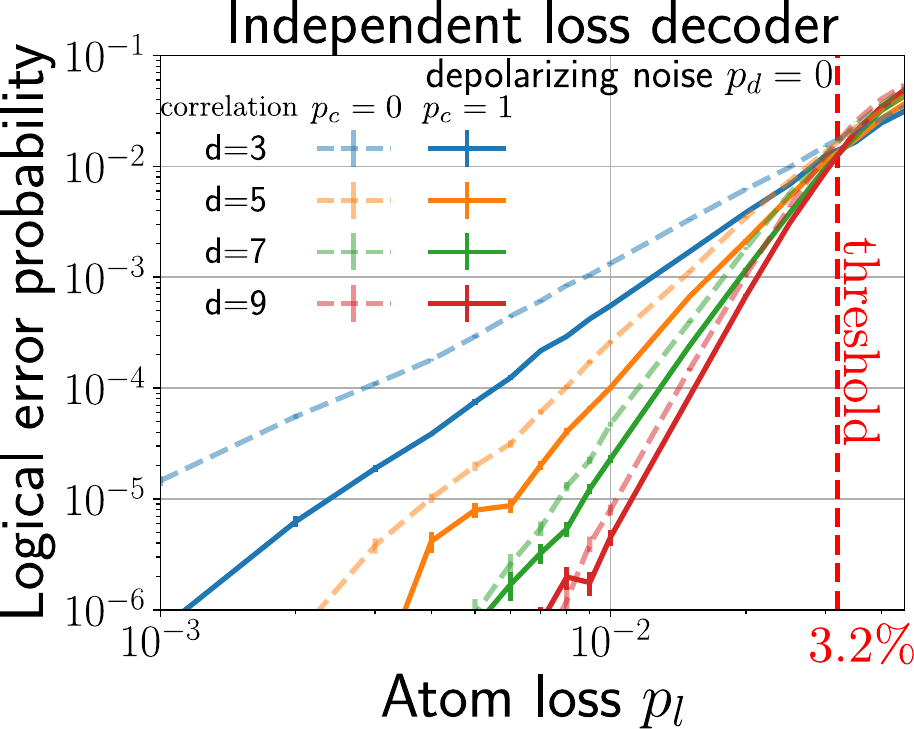}
    \caption{Logical error probability normalized by the number of rounds at vanishing depolarizing noise, $p_d = 0$, as a function of the atom loss probability $p_l$ for code distances $d = 3, 5, 7, 9$ with $d$ cycles of stabilizer measurements obtained by employing the independent-loss decoder of~\cite{perrin2025}. Solid lines correspond to results obtained for a fully correlated loss model $p_c=1$, while dashed lines show the logical error probabilities in the independent-loss regime. The dashed red vertical line indicates the independent-loss decoder threshold at $p_l = 3.2\%$ for both loss models.  At least $10^4$ shots were used to estimate the logical error probabilities, with up to $\sim10^6$ shots employed for the lowest error rates.}
    \label{fig:indpt_decoder}
\end{figure}

Figure~\ref{fig:indpt_decoder} benchmarks the independent-loss decoder introduced in~\cite{perrin2025} (summarized in Sec.~\ref{sec:independent}) at vanishing depolarizing noise ($p_d=0$) under fully correlated losses (solid lines) and compares it to the uncorrelated case (dashed lines). Interestingly, although fully correlated losses double the number of lost atoms relative to the uncorrelated model, the threshold remains unchanged at $3.2\%$, and the logical performance is slightly improved in the fully correlated regime.

In the independent-loss model, whenever a loss occurs, the remaining atom is subjected to a strong Pauli noise channel. As a result, this surviving qubit becomes effectively unusable for error correction during that round. In this sense, we argue that it is preferable for the second atom to be lost as well, since the loss is then heralded by the LDU and can be treated as a delayed erasure channel during decoding.

Note that for comparing the results of Fig.~\ref{fig:indpt_decoder} with the independent-loss model considered in~\cite{perrin2025}, the loss probability per gate used in the present work must be divided by a factor of two to match the loss probability per gate and per atom defined in~\cite{perrin2025}. From Fig.~\ref{fig:indpt_decoder}, we infer a threshold of \( 1.6\% \) for this rescaled quantity, which is lower than the threshold reported in~\cite{perrin2025}. This difference originates from the more realistic Pauli noise model adopted in the present work for the remaining atom following a loss event (see App.~\ref{ap:noise remaining}). In contrast, Ref.~\cite{perrin2025} either assumes no additional noise on the remaining atom, leading to a threshold of \( 2.6\% \), or considers a fully biased \( Z \)-error channel, yielding a threshold of \( 2.1\% \).

Nevertheless, it may be possible to tune the coefficients of the Pauli noise channel acting on the remaining atom through a careful encoding of the qubit subspace within the electronic level structure of the neutral atom. In particular, if decay from the Rydberg level back to the \( \ket{0} \) state is forbidden, the resulting noise channel reduces to a fully biased \( Z \)-error channel. In this scenario, the threshold at \( p_c = 0 \) is expected to increase to approximately \( 2\times2.1\% =4.2\% \), and increasing the correlation parameter \( p_c \) at fixed \( p_l \) will decrease the logical performance.\\

\begin{figure}[t!]
    \centering
    \includegraphics[width=\linewidth]{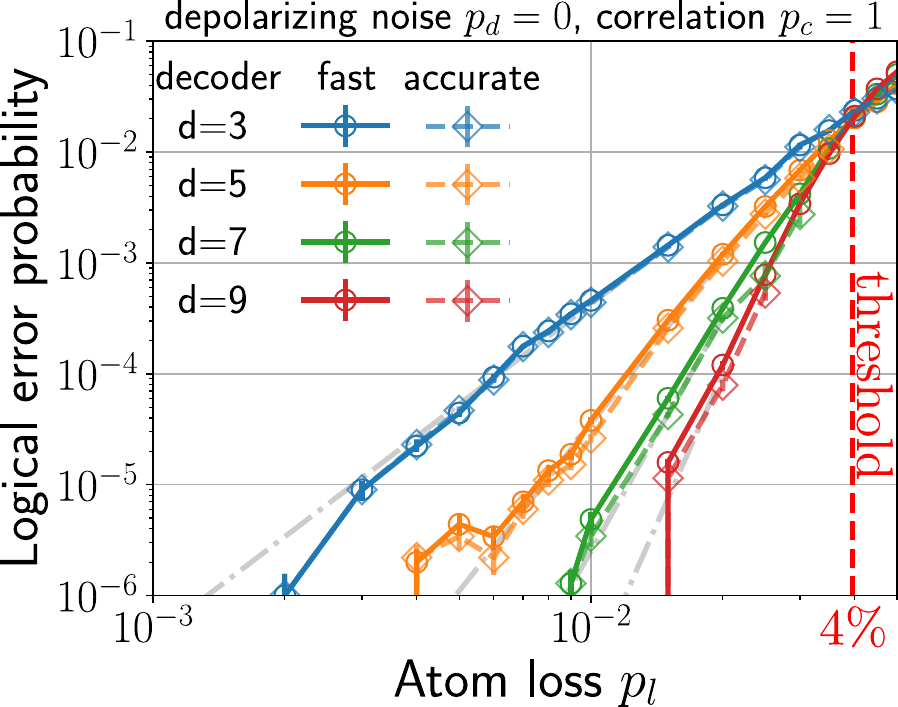}\\
    \caption{Logical error probability normalized by the number of rounds at vanishing depolarizing noise, $p_d = 0$, as a function of the fully correlated loss probability $p_l$ ($p_c = 1$), for code distances $d = 3, 5, 7, 9$ with $d$ cycles of stabilizer measurements. Solid lines with circular markers correspond to results obtained using the fast decoder, while dashed lines show the logical error probabilities decoded with the accurate decoder. The dashed red vertical line indicates the fast decoder threshold at $p_l = 4\%$. Semi-transparent gray dash-dotted curves show fits of power-law $d$ with respect to the fast decoder data. At least $10^4$ shots were used to estimate the logical error probabilities, with up to $10^6$ shots employed for the lowest error rates.}
    \label{fig:correlated_loss_no_depo}
\end{figure}

In Fig.~\ref{fig:correlated_loss_no_depo}, we present the performance of the two decoders introduced in Sec.~\ref{sec:accurate} and~\ref{sec:fast}, namely the accurate and the fast decoder, in the absence of depolarizing noise (\( p_d = 0 \)) and for a fully correlated loss model (\( p_c = 1 \)). We vary the loss probability in the range \( 0.001 \leq p_l \leq 0.045 \) and observe a threshold at $4\% $ for the fast decoder. The dash-dotted gray curves correspond to power-law fits with exponent $d$ of the fast decoder logical error probabilities (solid curves), consistent with the expected scaling for an erasure-like channel~\cite{nielsen2010}.

For the accurate decoder, simulations at large loss probabilities are computationally prohibitive, as the size of the loss graphs increases significantly, leading to an exponential growth of the runtime of the \( k \)-matching algorithm. At lower loss probabilities, the accurate decoder (dashed curves) provides only a marginal improvement over the fast decoder, while incurring a substantially larger decoding latency, rendering it impractical for real-time decoding. As a result, in the remainder of this work, we benchmark exclusively the performance of the fast decoder.\\ 

\begin{figure}[t!]
    \centering
    \includegraphics[width=\linewidth]{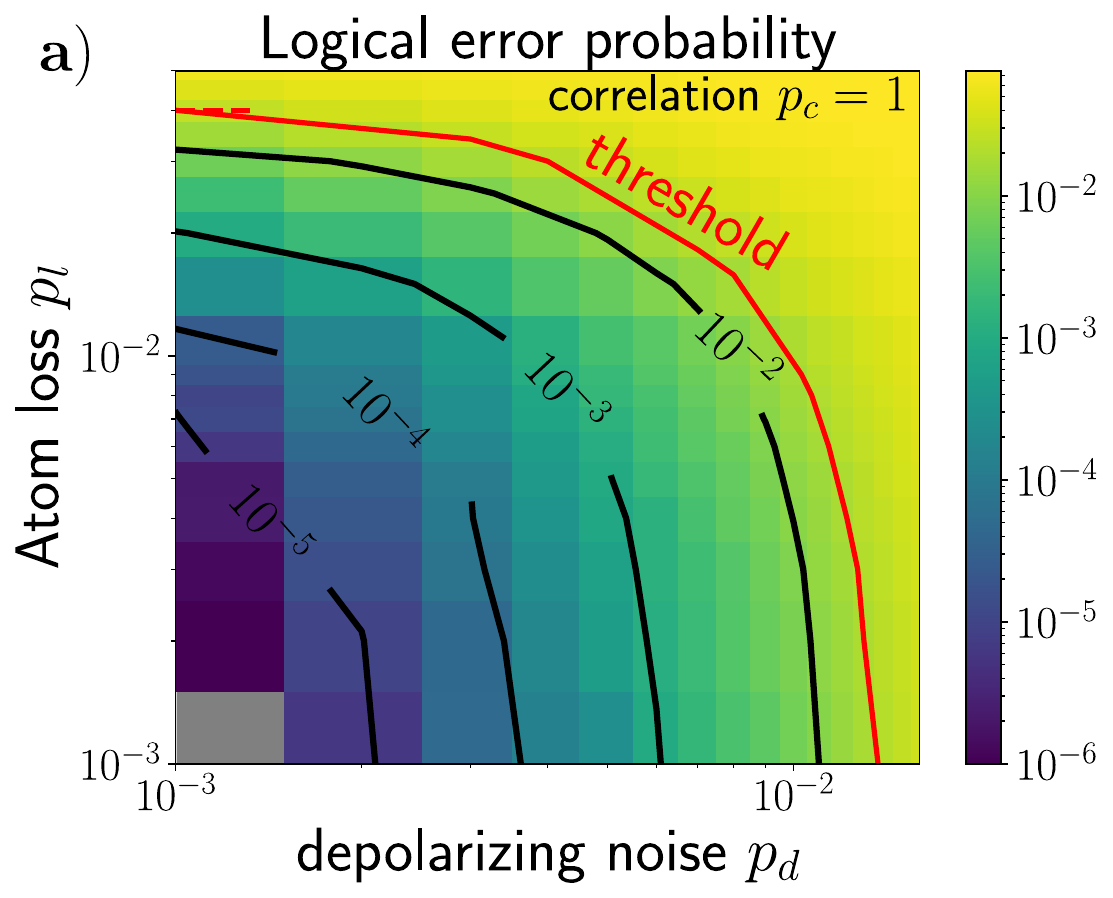}\\
    \includegraphics[width=\linewidth]{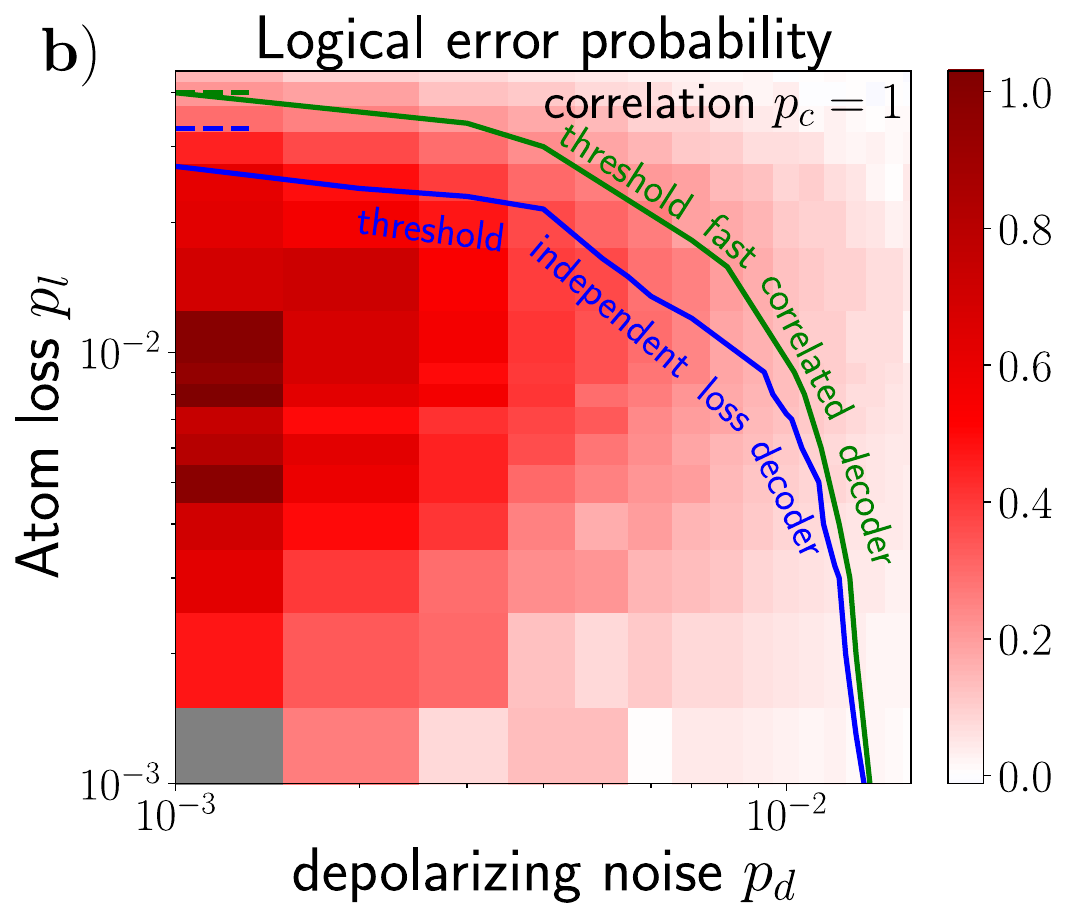}
    \caption{{\bf a)} Logical error probability normalized by the number of rounds
for a surface code of distance $d=9$ and $d$ rounds of stabilizer measurements as a function of the fully correlated loss probability $p_l$ ($p_c=1$) and the depolarizing error probability $p_d$. At least $10^4$ shots were used to estimate the logical error probabilities, with up to $10^6$ shots employed for the lowest error rates. The solid red line marks the error threshold while the solid black lines show curves of constant logical error probability.
{\bf b)} Plot showing the gain $G$ of the fast correlated decoder compared to the independent loss decoder of~\cite{perrin2025} as a function of the loss probability $p_l$ and the depolarizing noise error probability $p_d$ for a surface code of
distance $d=9$. The gray region indicates no errors were found for the fast correlated decoder.}
    \label{fig:correlated loss and depo}
\end{figure}

In Fig.~\ref{fig:correlated loss and depo}{\bf a}, we show the logical error probability for a distance code $d=9$ obtained using the fast decoder at various depolarizing noise probabilities $0.001\leq p_d \leq 0.016$ and loss probabilities $0.001\leq p_d \leq 0.05$ for a fully correlated loss model $p_c =1$. We indicate the threshold by a solid red line below which increasing the code distance, decreases exponentially the logical error. The plot illustrates the benefit of treating correlated atom loss distinctly from depolarizing noise. As an example, to achieve a logical error rate of $10^{-4}$, it is sufficient to tolerate correlated atom loss up to $1\%$ per CZ gate when the depolarizing noise is fixed at $10^{-3}$. Conversely, when correlated atom loss is fixed at $10^{-3}$, the same logical error rate can be reached with depolarizing noise as large as $3.5 \times 10^{-3}$.

We further emphasize the importance of employing a decoding strategy that is capable of leveraging loss correlations. To quantify this advantage, we introduce the gain \( G \), defined as
\begin{equation}
    G = \log_{10}\!\left(\frac{\varepsilon_r^{\text{indpt.}}}{\varepsilon_r^{\text{corr.}}}\right),
    \label{eq:gain}
\end{equation}
where \( \varepsilon_r^{\text{indpt.}} \) denotes the logical error probability per round obtained with a decoder that assumes independent loss events, and \( \varepsilon_r^{\text{corr.}} \) is the logical error probability per round achieved using the fast decoder, which accounts for loss correlations. The gain \( G \) is positive whenever the fast decoder outperforms the independent-loss decoder; for instance, \( G = 1 \) corresponds to a one-order-of-magnitude improvement in the logical error.

With the same noise channel as in Fig.~\ref{fig:correlated loss and depo}{\bf a} (i.e. a fully correlated loss channel $p_c=1$, and varying the loss probability $p_l$ as well as the depolarizing probability $p_d$), we plot in Fig.~\ref{fig:correlated loss and depo}{\bf b} the gain. In the regime where correlated atom loss are the dominant source of noise, we show that it is advantageous to consider the fast correlated loss decoder. In particular, the threshold discrepancy between both decoders increases as the ratio of correlated atom loss to depolarizing noise increases and the logical error performance may improve up to one order of magnitude.  

\subsection{Partial correlated loss}

So far, simulations assumed atoms are always lost in pairs—an idealized scenario that maximizes decoding performance but is not fully realistic. As discussed in Sec.~\ref{sec:accurate}, the decoder can be extended to handle partially correlated losses.\\

\begin{figure}[!!h]
    \centering
    \includegraphics[width=\linewidth]{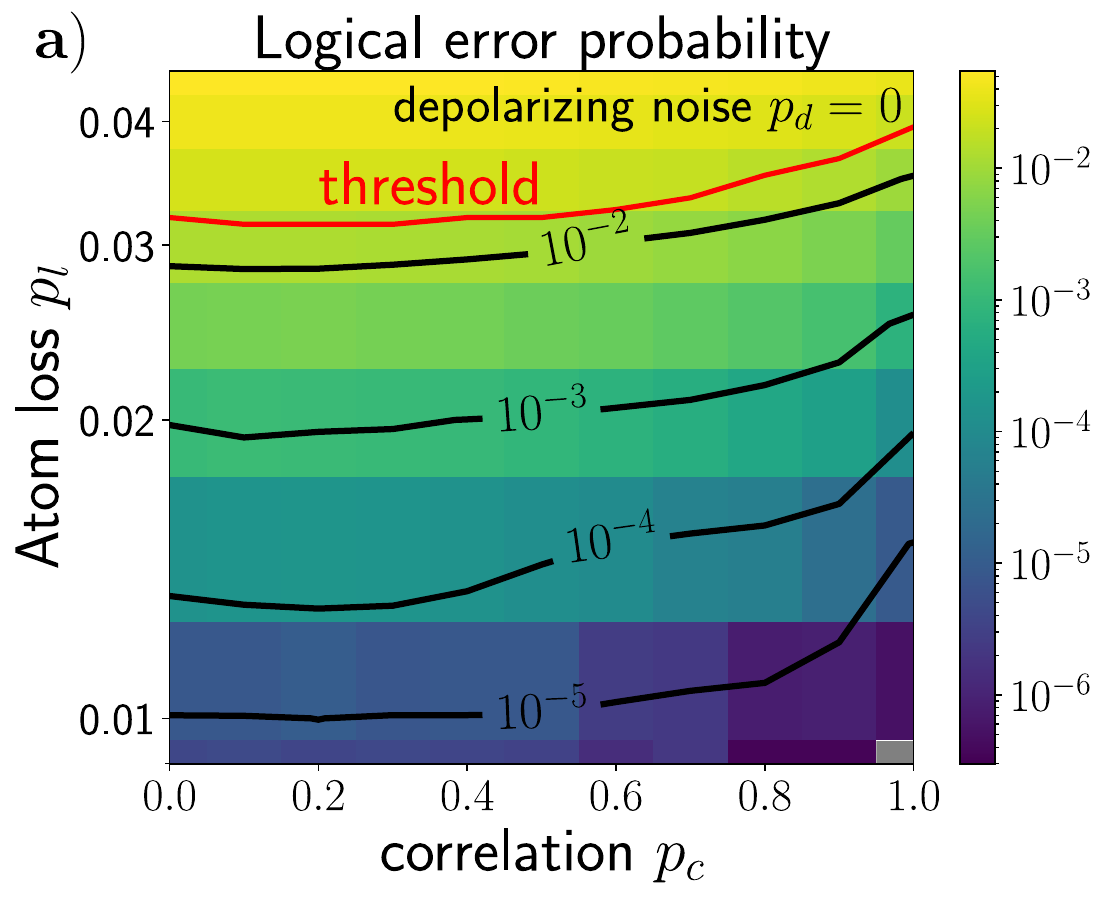}
    \includegraphics[width = \linewidth]{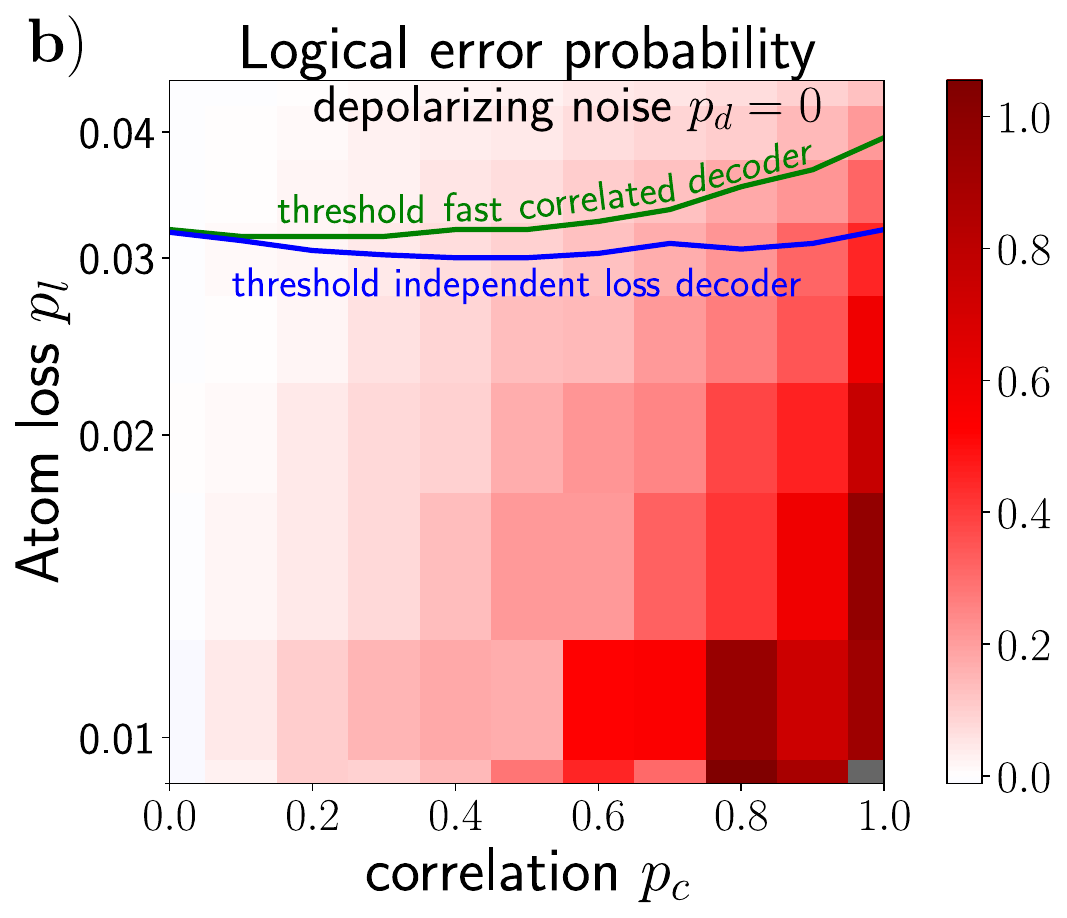}
    \caption{{\bf a)} Logical error probability normalized by the number of rounds
for a surface code of distance $d=9$ and $d$ rounds of stabilizer measurements at vanish depolarizing noise $p_d=0$ as a function of the atom loss probability $p_l$ and the correlation probability $p_c$. At least $10^4$ shots were used to estimate the logical error probabilities, with up to $10^6$ shots employed for the lowest error rates. The solid red line marks the error threshold while the solid black lines show curves of constant logical error probability.
{\bf b)} Plot showing the gain $G$ of the fast correlated decoder compared to the independent loss decoder of~\cite{perrin2025} as a function of the atom loss probability $p_l$ and the correlation probability $p_c$ without depolarizing noise for a surface code of
distance $d=9$. The gray region indicates no errors were found for the fast correlated decoder.}
    \label{fig:partial_correlated_loss}
\end{figure}

In Fig.~\ref{fig:partial_correlated_loss}{\bf a}, we present the logical error probability per round for a surface code of distance ($d = 9 $) as a function of the loss probability \( p_l \) and the correlation parameter \( 0 \leq p_c \leq 1 \), while keeping the depolarizing noise probability fixed at \( p_d = 0 \). Taking advantage of correlations in atom loss, we observe a clear improvement in logical performance as the correlation strength increases. In particular, the threshold rises from $3.2\%$ to $4\%$. More precisely, for correlations 
$p_c\leq0.5$, the logical performance is preserved despite an increased marginal loss rate (as already noticed for the independent-loss decoder), whereas for $p_c>0.5$ the decoder accuracy improves significantly.

Figure~\ref{fig:partial_correlated_loss}{\bf b} shows the relative gain of using a correlated decoder over an independent-loss decoder. At vanishing correlation, both decoders perform equally well. As correlation increases, the independent-loss decoder threshold remains roughly constant, while the correlated decoder threshold rises. The gain is maximal at full correlation, $p_c=1$.

\section{Conclusion}
In conclusion, we benchmarked the effect of correlated atom loss on the surface code with teleportation-based LDUs. Remarkably, even when correlations are ignored at decoding, logical performance can improve despite the total number of losses being doubled. This arises because LDUs herald both lost atoms in correlated events, whereas uncorrelated losses contaminate neighboring qubits with strong Pauli noise.

Incorporating correlations into the decoder further boosts performance, reducing the logical error rate by up to an order of magnitude for a distance-$9$ surface code and raising the threshold from 3.2\% to 4\%. This is achieved by constructing a loss graph and updating edge probabilities based on the observed syndrome — a procedure that is highly parallelizable and compatible with real-time decoding. The approach can also be adapted to partially correlated losses.

Our work highlights the value of architecture-specific loss models. For neutral-atom platforms, it is crucial to quantify the balance between correlated and uncorrelated losses during CZ gates, while also accounting for other channels such as background-gas collisions and atom transport. Leakage can similarly be treated as loss if it decays between operations, but faster gate cycles may require modeling residual leakage explicitly to minimize clock time.

Finally, atom-loss probabilities are in principle state-dependent (see Appendix~\ref{ap:state dependent}), motivating more accurate simulations and corresponding decoder extensions as a promising direction for future work.

\bibliography{atom_loss_initials}

@article{acharya2023,
  title = {Suppressing {{Quantum Errors}} by {{Scaling}} a {{Surface Code Logical Qubit}}},
author = {Acharya, R. and {al.}},
  year = 2023,
  month = feb,
  journal = {Nature},
  volume = {614},
  number = {7949},
  pages = {676--681},
  publisher = {Nature Publishing Group},
  issn = {1476-4687},
  doi = {10.1038/s41586-022-05434-1},
  url = {https://www.nature.com/articles/s41586-022-05434-1},
  urldate = {2024-09-24},
  copyright = {2023 The Author(s)}
}

@article{acharya2025,
  title = {Quantum {{Error Correction}} below the {{Surface Code Threshold}}},
author = {Acharya, R. and {al.}},
  year = 2025,
  month = feb,
  journal = {Nature},
  volume = {638},
  number = {8052},
  pages = {920--926},
  publisher = {Nature Publishing Group},
  issn = {1476-4687},
  doi = {10.1038/s41586-024-08449-y},
  url = {https://www.nature.com/articles/s41586-024-08449-y},
  urldate = {2025-03-07},
  copyright = {2024 The Author(s)}
}

@article{anand2024,
  title = {A {{Dual-Species Rydberg Array}}},
author = {Anand, S. and Bradley, C. E. and White, R. and Ramesh, V. and Singh, K. and Bernien, H.},
  year = 2024,
  month = nov,
  journal = {Nature Physics},
  volume = {20},
  number = {11},
  pages = {1744--1750},
  publisher = {Nature Publishing Group},
  issn = {1745-2481},
  doi = {10.1038/s41567-024-02638-2},
  url = {https://www.nature.com/articles/s41567-024-02638-2},
  urldate = {2024-12-09},
  copyright = {2024 The Author(s)}
}

@article{baranes2026,
  title = {Leveraging {{Qubit Loss Detection}} in {{Fault-Tolerant Quantum Algorithms}}},
author = {Baranes, G. and Cain, M. and Ataides, J. P. B. and Bluvstein, D. and Sinclair, J. and Vuleti{\'c}, V. and Zhou, H. and Lukin, M. D.},
  year = 2026,
  month = jan,
  journal = {Physical Review X},
  volume = {16},
  number = {1},
  pages = {011002},
  publisher = {American Physical Society},
  doi = {10.1103/ycwc-3myc},
  url = {https://link.aps.org/doi/10.1103/ycwc-3myc},
  urldate = {2026-03-23}
}

@article{bluvstein2022,
  title = {A {{Quantum Processor Based}} on {{Coherent Transport}} of {{Entangled Atom Arrays}}},
author = {Bluvstein, D. and {al.}},
  year = 2022,
  month = apr,
  journal = {Nature},
  volume = {604},
  number = {7906},
  pages = {451--456},
  publisher = {Nature Publishing Group},
  issn = {1476-4687},
  doi = {10.1038/s41586-022-04592-6},
  url = {https://www.nature.com/articles/s41586-022-04592-6},
  urldate = {2024-09-23},
  copyright = {2022 The Author(s)}
}

@article{bluvstein2024,
  title = {Logical {{Quantum Processor Based}} on {{Reconfigurable Atom Arrays}}},
author = {Bluvstein, D. and {al.}},
  year = 2024,
  month = feb,
  journal = {Nature},
  volume = {626},
  number = {7997},
  pages = {58--65},
  publisher = {Nature Publishing Group},
  issn = {1476-4687},
  doi = {10.1038/s41586-023-06927-3},
  url = {https://www.nature.com/articles/s41586-023-06927-3},
  urldate = {2024-09-16},
  copyright = {2023 The Author(s)}
}

@article{bravyi2024,
  title = {High-{{Threshold}} and {{Low-Overhead Fault-Tolerant Quantum Memory}}},
author = {Bravyi, S. and Cross, A. W. and Gambetta, J. M. and Maslov, D. and Rall, P. and Yoder, T. J.},
  year = 2024,
  month = mar,
  journal = {Nature},
  volume = {627},
  number = {8005},
  pages = {778--782},
  publisher = {Nature Publishing Group},
  issn = {1476-4687},
  doi = {10.1038/s41586-024-07107-7},
  url = {https://www.nature.com/articles/s41586-024-07107-7},
  urldate = {2025-03-10},
  copyright = {2024 The Author(s)}
}

@article{browaeys2020,
  title = {Many-{{Body Physics}} with {{Individually Controlled Rydberg Atoms}}},
author = {Browaeys, A. and Lahaye, T.},
  year = 2020,
  month = feb,
  journal = {Nature Physics},
  volume = {16},
  number = {2},
  pages = {132--142},
  publisher = {Nature Publishing Group},
  issn = {1745-2481},
  doi = {10.1038/s41567-019-0733-z},
  url = {https://www.nature.com/articles/s41567-019-0733-z},
  urldate = {2024-09-23},
  copyright = {2020 Springer Nature Limited}
}

@article{cao2024,
  title = {Multi-{{Qubit Gates}} and {{Schr\"odinger Cat States}} in an {{Optical Clock}}},
author = {Cao, A. and {al.}},
  year = 2024,
  month = oct,
  journal = {Nature},
  volume = {634},
  number = {8033},
  pages = {315--320},
  publisher = {Nature Publishing Group},
  issn = {1476-4687},
  doi = {10.1038/s41586-024-07913-z},
  url = {https://www.nature.com/articles/s41586-024-07913-z},
  urldate = {2024-12-09},
  copyright = {2024 This is a U.S. Government work and not under copyright protection in the US; foreign copyright protection may apply}
}

@article{chang2025,
  title = {Surface {{Code}} with {{Imperfect Erasure Checks}}},
author = {Chang, K. and Singh, S. and Claes, J. and Sahay, K. and Teoh, J. and Puri, S.},
  year = 2025,
  month = dec,
  journal = {PRX Quantum},
  volume = {6},
  number = {4},
  pages = {040355},
  publisher = {American Physical Society},
  doi = {10.1103/d1v7-nctj},
  url = {https://link.aps.org/doi/10.1103/d1v7-nctj},
  urldate = {2026-03-23}
}

@article{chow2024,
  title = {Circuit-{{Based Leakage-to-Erasure Conversion}} in a {{Neutral-Atom Quantum Processor}}},
author = {Chow, M. N. H. and Buchemmavari, V. and Omanakuttan, S. and Little, B. J. and Pandey, S. and Deutsch, I. H. and Jau, Y.},
  year = 2024,
  month = dec,
  journal = {PRX Quantum},
  volume = {5},
  number = {4},
  pages = {040343},
  publisher = {American Physical Society},
  doi = {10.1103/PRXQuantum.5.040343},
  url = {https://link.aps.org/doi/10.1103/PRXQuantum.5.040343},
  urldate = {2025-03-07}
}

@article{cong2022,
  title = {Hardware-{{Efficient}}, {{Fault-Tolerant Quantum Computation}} with {{Rydberg Atoms}}},
author = {Cong, I. and Levine, H. and Keesling, A. and Bluvstein, D. and Wang, S. and Lukin, M. D.},
  year = 2022,
  month = jun,
  journal = {Physical Review X},
  volume = {12},
  number = {2},
  pages = {021049},
  issn = {2160-3308},
  doi = {10.1103/PhysRevX.12.021049},
  url = {https://link.aps.org/doi/10.1103/PhysRevX.12.021049},
  urldate = {2024-09-23}
}

@misc{dasilva2024,
  title = {Demonstration of {{Logical Qubits}} and {{Repeated Error Correction}} with {{Better-than-Physical Error Rates}}},
author = {{da Silva}, M. P. and {al.}},
  year = 2024,
  month = apr,
  eprint = {2404.02280},
  primaryclass = {quant-ph},
  publisher = {arXiv},
  doi = {10.48550/arXiv.2404.02280},
  url = {http://arxiv.org/abs/2404.02280},
  urldate = {2024-09-16},
  archiveprefix = {arXiv}
}

@article{egan2021,
  title = {Fault-{{Tolerant Control}} of an {{Error-Corrected Qubit}}},
author = {Egan, L. and {al.}},
  year = 2021,
  month = oct,
  journal = {Nature},
  volume = {598},
  number = {7880},
  pages = {281--286},
  publisher = {Nature Publishing Group},
  issn = {1476-4687},
  doi = {10.1038/s41586-021-03928-y},
  url = {https://www.nature.com/articles/s41586-021-03928-y},
  urldate = {2024-09-23},
  copyright = {2021 The Author(s), under exclusive licence to Springer Nature Limited}
}

@misc{eisert2025,
  title = {Mind the Gaps: {{The}} Fraught Road to Quantum Advantage},
  shorttitle = {Mind the Gaps},
author = {Eisert, J. and Preskill, J.},
  year = 2025,
  month = nov,
  number = {arXiv:2510.19928},
  eprint = {2510.19928},
  primaryclass = {quant-ph},
  publisher = {arXiv},
  doi = {10.48550/arXiv.2510.19928},
  url = {http://arxiv.org/abs/2510.19928},
  urldate = {2026-03-10},
  archiveprefix = {arXiv}
}

@article{evered2023,
  title = {High-{{Fidelity Parallel Entangling Gates}} on a {{Neutral-Atom Quantum Computer}}},
author = {Evered, S. J. and {al.}},
  year = 2023,
  month = oct,
  journal = {Nature},
  volume = {622},
  number = {7982},
  pages = {268--272},
  publisher = {Nature Publishing Group},
  issn = {1476-4687},
  doi = {10.1038/s41586-023-06481-y},
  url = {https://www.nature.com/articles/s41586-023-06481-y},
  urldate = {2024-09-16},
  copyright = {2023 The Author(s)}
}

@article{finkelstein2024,
  title = {Universal {{Quantum Operations}} and {{Ancilla-Based Read-out}} for {{Tweezer Clocks}}},
author = {Finkelstein, R. and Tsai, R. B. and Sun, X. and Scholl, P. and Direkci, S. and Gefen, T. and Choi, J. and Shaw, A. L. and Endres, M.},
  year = 2024,
  month = oct,
  journal = {Nature},
  volume = {634},
  number = {8033},
  pages = {321--327},
  publisher = {Nature Publishing Group},
  issn = {1476-4687},
  doi = {10.1038/s41586-024-08005-8},
  url = {https://www.nature.com/articles/s41586-024-08005-8},
  urldate = {2024-12-09},
  copyright = {2024 The Author(s)}
}

@article{fowler2012,
  title = {Surface {{Codes}}: {{Towards Practical Large-Scale Quantum Computation}}},
  shorttitle = {Surface {{Codes}}},
author = {Fowler, A. G. and Mariantoni, M. and Martinis, J. M. and Cleland, A. N.},
  year = 2012,
  month = sep,
  journal = {Physical Review A},
  volume = {86},
  number = {3},
  pages = {032324},
  publisher = {American Physical Society},
  doi = {10.1103/PhysRevA.86.032324},
  url = {https://link.aps.org/doi/10.1103/PhysRevA.86.032324},
  urldate = {2024-09-23}
}

@article{gidney2021,
  title = {Stim: {{A Fast Stabilizer Circuit Simulator}}},
  shorttitle = {Stim},
author = {Gidney, C.},
  year = 2021,
  month = jul,
  journal = {Quantum},
  volume = {5},
  pages = {497},
  publisher = {Verein zur F\"orderung des Open Access Publizierens in den Quantenwissenschaften},
  doi = {10.22331/q-2021-07-06-497},
  url = {https://quantum-journal.org/papers/q-2021-07-06-497/},
  urldate = {2024-09-23}
}

@misc{gottesman1997,
  title = {Stabilizer {{Codes}} and {{Quantum Error Correction}}},
author = {Gottesman, D.},
  year = 1997,
  month = may,
  number = {arXiv:quant-ph/9705052},
  publisher = {arXiv},
  doi = {10.48550/arXiv.quant-ph/9705052},
  url = {http://arxiv.org/abs/quant-ph/9705052},
  urldate = {2024-09-16}
}

@article{graham2019,
  title = {Rydberg-{{Mediated Entanglement}} in a {{Two-Dimensional Neutral Atom Qubit Array}}},
author = {Graham, T. M. and Kwon, M. and Grinkemeyer, B. and Marra, Z. and Jiang, X. and Lichtman, M. T. and Sun, Y. and Ebert, M. and Saffman, M.},
  year = 2019,
  month = dec,
  journal = {Physical Review Letters},
  volume = {123},
  number = {23},
  pages = {230501},
  publisher = {American Physical Society},
  doi = {10.1103/PhysRevLett.123.230501},
  url = {https://link.aps.org/doi/10.1103/PhysRevLett.123.230501},
  urldate = {2026-03-10}
}

@article{graham2022,
  title = {Multi-{{Qubit Entanglement}} and {{Algorithms}} on a {{Neutral-Atom Quantum Computer}}},
author = {Graham, T. M. and {al.}},
  year = 2022,
  month = apr,
  journal = {Nature},
  volume = {604},
  number = {7906},
  pages = {457--462},
  publisher = {Nature Publishing Group},
  issn = {1476-4687},
  doi = {10.1038/s41586-022-04603-6},
  url = {https://www.nature.com/articles/s41586-022-04603-6},
  urldate = {2024-09-23},
  copyright = {2022 The Author(s), under exclusive licence to Springer Nature Limited}
}

@misc{gu2024,
  title = {Optimizing {{Quantum Error Correction Protocols}} with {{Erasure Qubits}}},
author = {Gu, S. and Vaknin, Y. and Retzker, A. and Kubica, A.},
  year = 2024,
  month = aug,
  eprint = {2408.00829},
  publisher = {arXiv},
  doi = {10.48550/arXiv.2408.00829},
  url = {http://arxiv.org/abs/2408.00829},
  urldate = {2025-03-10},
  archiveprefix = {arXiv}
}

@article{gu2025,
  title = {Fault-{{Tolerant Quantum Architectures Based}} on {{Erasure Qubits}}},
author = {Gu, S. and Retzker, A. and Kubica, A.},
  year = 2025,
  month = mar,
  journal = {Physical Review Research},
  volume = {7},
  number = {1},
  pages = {013249},
  publisher = {American Physical Society},
  doi = {10.1103/PhysRevResearch.7.013249},
  url = {https://link.aps.org/doi/10.1103/PhysRevResearch.7.013249},
  urldate = {2025-03-10}
}

@article{hashim2021,
  title = {Randomized {{Compiling}} for {{Scalable Quantum Computing}} on a {{Noisy Superconducting Quantum Processor}}},
author = {Hashim, A. and {al.}},
  year = 2021,
  month = nov,
  journal = {Physical Review X},
  volume = {11},
  number = {4},
  pages = {041039},
  publisher = {American Physical Society},
  doi = {10.1103/PhysRevX.11.041039},
  url = {https://link.aps.org/doi/10.1103/PhysRevX.11.041039},
  urldate = {2024-09-23}
}

@article{henriet2020,
  title = {Quantum {{Computing}} with {{Neutral Atoms}}},
author = {Henriet, L. and Beguin, L. and Signoles, A. and Lahaye, T. and Browaeys, A. and Reymond, G. and Jurczak, C.},
  year = 2020,
  month = sep,
  journal = {Quantum},
  volume = {4},
  pages = {327},
  publisher = {Verein zur F\"orderung des Open Access Publizierens in den Quantenwissenschaften},
  doi = {10.22331/q-2020-09-21-327},
  url = {https://quantum-journal.org/papers/q-2020-09-21-327/},
  urldate = {2024-09-23}
}

@misc{higgott2021,
  title = {{{PyMatching}}: {{A Python Package}} for {{Decoding Quantum Codes}} with {{Minimum-Weight Perfect Matching}}},
  shorttitle = {{{PyMatching}}},
author = {Higgott, O.},
  year = 2021,
  month = jul,
  eprint = {2105.13082},
  publisher = {arXiv},
  doi = {10.48550/arXiv.2105.13082},
  url = {http://arxiv.org/abs/2105.13082},
  urldate = {2024-09-23},
  archiveprefix = {arXiv}
}

@article{higgott2025,
  title = {Sparse {{Blossom}}: {{Correcting}} a {{Million Errors}} per {{Core Second}} with {{Minimum-Weight Matching}}},
  shorttitle = {Sparse {{Blossom}}},
author = {Higgott, O. and Gidney, C.},
  year = 2025,
  month = jan,
  journal = {Quantum},
  volume = {9},
  pages = {1600},
  issn = {2521-327X},
  doi = {10.22331/q-2025-01-20-1600},
  url = {http://arxiv.org/abs/2303.15933},
  urldate = {2025-01-23}
}

@article{jaksch2000,
  title = {Fast {{Quantum Gates}} for {{Neutral Atoms}}},
author = {Jaksch, D. and Cirac, J. I. and Zoller, P. and Rolston, S. L. and C{\^o}t{\'e}, R. and Lukin, M. D.},
  year = 2000,
  month = sep,
  journal = {Physical Review Letters},
  volume = {85},
  number = {10},
  pages = {2208--2211},
  issn = {0031-9007, 1079-7114},
  doi = {10.1103/PhysRevLett.85.2208},
  url = {https://link.aps.org/doi/10.1103/PhysRevLett.85.2208},
  urldate = {2024-12-06},
  copyright = {http://link.aps.org/licenses/aps-default-license}
}

@article{jandura2022,
  title = {Time-{{Optimal Two-}} and {{Three-Qubit Gates}} for {{Rydberg Atoms}}},
author = {Jandura, S. and Pupillo, G.},
  year = 2022,
  month = may,
  journal = {Quantum},
  volume = {6},
  pages = {712},
  publisher = {Verein zur F\"orderung des Open Access Publizierens in den Quantenwissenschaften},
  doi = {10.22331/q-2022-05-13-712},
  url = {https://quantum-journal.org/papers/q-2022-05-13-712/},
  urldate = {2024-09-23}
}

@misc{jandura2026,
  title = {Surface {{Code Stabilizer Measurements}} for {{Rydberg Atoms}}},
author = {Jandura, S. and Pecorari, L. and Pupillo, G.},
  year = 2026,
  month = mar,
  number = {arXiv:2405.16621},
  eprint = {2405.16621},
  primaryclass = {quant-ph},
  publisher = {arXiv},
  doi = {10.48550/arXiv.2405.16621},
  url = {http://arxiv.org/abs/2405.16621},
  urldate = {2026-03-23},
  archiveprefix = {arXiv}
}

@article{kern2005,
  title = {Quantum {{Error Correction}} of {{Coherent Errors}} by {{Randomization}}},
author = {Kern, O. and Alber, G. and Shepelyansky, D. L.},
  year = 2005,
  month = jan,
  journal = {The European Physical Journal D - Atomic, Molecular, Optical and Plasma Physics},
  volume = {32},
  number = {1},
  pages = {153--156},
  issn = {1434-6079},
  doi = {10.1140/epjd/e2004-00196-9},
  url = {https://doi.org/10.1140/epjd/e2004-00196-9},
  urldate = {2024-09-23}
}

@article{knill2005,
  title = {Quantum {{Computing}} with {{Realistically Noisy Devices}}},
author = {Knill, E.},
  year = 2005,
  month = mar,
  journal = {Nature},
  volume = {434},
  number = {7029},
  pages = {39--44},
  publisher = {Nature Publishing Group},
  issn = {1476-4687},
  doi = {10.1038/nature03350},
  url = {https://www.nature.com/articles/nature03350},
  urldate = {2025-06-18},
  copyright = {2005 Macmillan Magazines Ltd.}
}

@article{kobayashi2026,
  title = {Erasure-{{Tolerance Scheme}} for the {{Surface Codes}} on {{Neutral Atom Quantum Computers}}},
author = {Kobayashi, F. and Nagayama, S.},
  year = 2026,
  journal = {IEEE Transactions on Quantum Engineering},
  volume = {7},
  pages = {1--13},
  issn = {2689-1808},
  doi = {10.1109/TQE.2025.3627918},
  url = {https://ieeexplore.ieee.org/document/11223708},
  urldate = {2026-03-23}
}

@article{kubica2023,
  title = {Erasure {{Qubits}}: {{Overcoming}} the \${{T}}\_1\$  Limit in {{Superconducting Circuits}}},
  shorttitle = {Erasure {{Qubits}}},
author = {Kubica, A. and Haim, A. and Vaknin, Y. and Levine, H. and Brand{\~a}o, F. and Retzker, A.},
  year = 2023,
  month = nov,
  journal = {Physical Review X},
  volume = {13},
  number = {4},
  pages = {041022},
  publisher = {American Physical Society},
  doi = {10.1103/PhysRevX.13.041022},
  url = {https://link.aps.org/doi/10.1103/PhysRevX.13.041022},
  urldate = {2025-03-10}
}

@misc{liu2026,
  title = {Achieving {{Optimal-Distance Atom-Loss Correction}} via {{Pauli Envelope}}},
author = {Liu, P. and Tan, S. J. S. and Huang, E. and Acar, U. A. and Zhou, H. and Zhao, C.},
  year = 2026,
  month = mar,
  number = {arXiv:2603.04156},
  eprint = {2603.04156},
  primaryclass = {quant-ph},
  publisher = {arXiv},
  doi = {10.48550/arXiv.2603.04156},
  url = {http://arxiv.org/abs/2603.04156},
  urldate = {2026-03-10},
  archiveprefix = {arXiv},
  langid = {english}
}

@article{ma2023,
  title = {High-{{Fidelity Gates}} and {{Mid-Circuit Erasure Conversion}} in an {{Atomic Qubit}}},
author = {Ma, S. and Liu, G. and Peng, P. and Zhang, B. and Jandura, S. and Claes, J. and Burgers, A. P. and Pupillo, G. and Puri, S. and Thompson, J. D.},
  year = 2023,
  month = oct,
  journal = {Nature},
  volume = {622},
  number = {7982},
  pages = {279--284},
  publisher = {Nature Publishing Group},
  issn = {1476-4687},
  doi = {10.1038/s41586-023-06438-1},
  url = {https://www.nature.com/articles/s41586-023-06438-1},
  urldate = {2024-09-23},
  copyright = {2023 The Author(s), under exclusive licence to Springer Nature Limited}
}

@article{morgado2021,
  title = {Quantum {{Simulation}} and {{Computing}} with {{Rydberg-interacting Qubits}}},
author = {Morgado, M. and Whitlock, S.},
  year = 2021,
  month = may,
  journal = {AVS Quantum Science},
  volume = {3},
  number = {2},
  pages = {023501},
  issn = {2639-0213},
  doi = {10.1116/5.0036562},
  url = {https://doi.org/10.1116/5.0036562},
  urldate = {2024-09-23}
}

@article{moses2023,
  title = {A {{Race-Track Trapped-Ion Quantum Processor}}},
author = {Moses, S. A. and {al.}},
  year = 2023,
  month = dec,
  journal = {Physical Review X},
  volume = {13},
  number = {4},
  pages = {041052},
  issn = {2160-3308},
  doi = {10.1103/PhysRevX.13.041052},
  url = {https://link.aps.org/doi/10.1103/PhysRevX.13.041052},
  urldate = {2024-09-23}
}

@misc{nielsen2010,
  title = {Quantum {{Computation}} and {{Quantum Information}}: 10th {{Anniversary Edition}}},
  shorttitle = {Quantum {{Computation}} and {{Quantum Information}}},
author = {Nielsen, M. A. and Chuang, I. L.},
  year = 2010,
  month = dec,
  journal = {Higher Education from Cambridge University Press},
  publisher = {Cambridge University Press},
  doi = {10.1017/CBO9780511976667},
  urldate = {2024-10-21},
  isbn = {9780511976667}
}

@article{niroula2024,
  title = {Quantum {{Sensing}} with {{Erasure Qubits}}},
author = {Niroula, P. and Dolde, J. and Zheng, X. and Bringewatt, J. and Ehrenberg, A. and Cox, K. C. and Thompson, J. and Gullans, M. J. and Kolkowitz, S. and Gorshkov, A. V.},
  year = 2024,
  month = aug,
  journal = {Physical Review Letters},
  volume = {133},
  number = {8},
  pages = {080801},
  publisher = {American Physical Society},
  doi = {10.1103/PhysRevLett.133.080801},
  url = {https://link.aps.org/doi/10.1103/PhysRevLett.133.080801},
  urldate = {2024-12-09}
}

@misc{omanakuttan2024,
  title = {Coherence {{Preserving Leakage Detection}} and {{Cooling}} in {{Alkaline Earth Atoms}}},
author = {Omanakuttan, S. and Buchemmavari, V. and Martin, M. J. and Deutsch, I. H.},
  year = 2024,
  month = oct,
  eprint = {2410.23430},
  publisher = {arXiv},
  doi = {10.48550/arXiv.2410.23430},
  url = {http://arxiv.org/abs/2410.23430},
  urldate = {2025-01-06},
  archiveprefix = {arXiv}
}

@article{pagano2022,
  title = {Error {{Budgeting}} for a {{Controlled-Phase Gate}} with {{Strontium-88 Rydberg Atoms}}},
author = {Pagano, A. and Weber, S. and Jaschke, D. and Pfau, T. and Meinert, F. and Montangero, S. and B{\"u}chler, H. P.},
  year = 2022,
  month = jul,
  journal = {Physical Review Research},
  volume = {4},
  number = {3},
  pages = {033019},
  publisher = {American Physical Society},
  doi = {10.1103/PhysRevResearch.4.033019},
  url = {https://link.aps.org/doi/10.1103/PhysRevResearch.4.033019},
  urldate = {2024-09-23}
}

@article{pecorari2025a,
  title = {High-{{Rate Quantum LDPC Codes}} for {{Long-Range-Connected Neutral Atom Registers}}},
author = {Pecorari, L. and Jandura, S. and Brennen, G. K. and Pupillo, G.},
  year = 2025,
  month = jan,
  journal = {Nature Communications},
  volume = {16},
  number = {1},
  pages = {1111},
  publisher = {Nature Publishing Group},
  issn = {2041-1723},
  doi = {10.1038/s41467-025-56255-5},
  url = {https://www.nature.com/articles/s41467-025-56255-5},
  urldate = {2025-02-06},
  copyright = {2025 The Author(s)}
}

@article{pecorari2025b,
  title = {Quantum Low-Density Parity-Check Codes for Erasure-Biased Atomic Quantum Processors},
author = {Pecorari, L. and Pupillo, G.},
  year = 2025,
  month = nov,
  journal = {Physical Review A},
  volume = {112},
  number = {5},
  pages = {052417},
  publisher = {American Physical Society},
  doi = {10.1103/mgkt-ctv8},
  url = {https://link.aps.org/doi/10.1103/mgkt-ctv8},
  urldate = {2026-03-23}
}

@article{perrin2025,
  title = {Quantum {{Error Correction}} Resilient against {{Atom Loss}}},
author = {Perrin, H. and Jandura, S. and Pupillo, G.},
  year = 2025,
  month = oct,
  journal = {Quantum},
  volume = {9},
  pages = {1884},
  publisher = {Verein zur F\"orderung des Open Access Publizierens in den Quantenwissenschaften},
  doi = {10.22331/q-2025-10-13-1884},
  url = {https://quantum-journal.org/papers/q-2025-10-13-1884/},
  urldate = {2026-03-23},
  langid = {british}
}

@article{postler2022,
  title = {Demonstration of {{Fault-Tolerant Universal Quantum Gate Operations}}},
author = {Postler, L. and {al.}},
  year = 2022,
  month = may,
  journal = {Nature},
  volume = {605},
  number = {7911},
  pages = {675--680},
  publisher = {Nature Publishing Group},
  issn = {1476-4687},
  doi = {10.1038/s41586-022-04721-1},
  url = {https://www.nature.com/articles/s41586-022-04721-1},
  urldate = {2024-09-16},
  copyright = {2022 The Author(s), under exclusive licence to Springer Nature Limited}
}

@incollection{preskill1998,
  title = {Fault-{{Tolerant Quantum Computation}}},
  booktitle = {Introduction to {{Quantum Computation}} and {{Information}}},
author = {Preskill, J.},
  year = 1998,
  month = oct,
  pages = {213--269},
  publisher = {WORLD SCIENTIFIC},
  doi = {10.1142/9789812385253_0008},
  url = {https://www.worldscientific.com/doi/abs/10.1142/9789812385253_0008},
  urldate = {2024-09-16},
  isbn = {978-981-02-3399-0}
}

@article{radnaev2025,
  title = {Universal {{Neutral-Atom Quantum Computer}} with {{Individual Optical Addressing}} and {{Nondestructive Readout}}},
author = {Radnaev, A. and al.},
  year = 2025,
  month = aug,
  journal = {PRX Quantum},
  volume = {6},
  number = {3},
  pages = {030334},
  publisher = {American Physical Society},
  doi = {10.1103/66s8-jj18},
  url = {https://link.aps.org/doi/10.1103/66s8-jj18},
  urldate = {2026-03-23}
}

@misc{reichardt2025,
  title = {Fault-Tolerant Quantum Computation with a Neutral Atom Processor},
author = {Reichardt, B. W. and al.},
  year = 2025,
  month = jun,
  number = {arXiv:2411.11822},
  eprint = {2411.11822},
  primaryclass = {quant-ph},
  publisher = {arXiv},
  doi = {10.48550/arXiv.2411.11822},
  url = {http://arxiv.org/abs/2411.11822},
  urldate = {2026-01-05},
  archiveprefix = {arXiv}
}

@article{ryan-anderson2021,
  title = {Realization of {{Real-Time Fault-Tolerant Quantum Error Correction}}},
author = {{Ryan-Anderson}, C. and {al.}},
  year = 2021,
  month = dec,
  journal = {Physical Review X},
  volume = {11},
  number = {4},
  pages = {041058},
  publisher = {American Physical Society},
  doi = {10.1103/PhysRevX.11.041058},
  url = {https://link.aps.org/doi/10.1103/PhysRevX.11.041058},
  urldate = {2024-09-16}
}

@article{saffman2010,
  title = {Quantum {{Information}} with {{Rydberg Atoms}}},
author = {Saffman, M. and Walker, T. G. and M{\o}lmer, K.},
  year = 2010,
  month = aug,
  journal = {Reviews of Modern Physics},
  volume = {82},
  number = {3},
  pages = {2313--2363},
  publisher = {American Physical Society},
  doi = {10.1103/RevModPhys.82.2313},
  url = {https://link.aps.org/doi/10.1103/RevModPhys.82.2313},
  urldate = {2024-09-23}
}

@article{scholl2023,
  title = {Erasure {{Conversion}} in a {{High-Fidelity Rydberg Quantum Simulator}}},
author = {Scholl, P. and Shaw, A. L. and Tsai, R. B. and Finkelstein, R. and Choi, J. and Endres, M.},
  year = 2023,
  month = oct,
  journal = {Nature},
  volume = {622},
  number = {7982},
  pages = {273--278},
  publisher = {Nature Publishing Group},
  issn = {1476-4687},
  doi = {10.1038/s41586-023-06516-4},
  url = {https://www.nature.com/articles/s41586-023-06516-4},
  urldate = {2024-09-23},
  copyright = {2023 The Author(s)}
}

@misc{senoo2025,
  title = {High-Fidelity Entanglement and Coherent Multi-Qubit Mapping in an Atom Array},
author = {Senoo, A. and Baumg{\"a}rtner, A. and Lis, J. W. and Vaidya, G. M. and Zeng, Z. and Giudici, G. and Pichler, H. and Kaufman, A. M.},
  year = 2025,
  month = jun,
  journal = {arXiv.org},
  eprint = {2506.13632},
  doi = {10.48550/arXiv.2506.13632},
  url = {https://arxiv.org/abs/2506.13632v2},
  urldate = {2026-03-23},
  archiveprefix = {arXiv},
  langid = {english}
}

@article{stricker2020,
  title = {Experimental {{Deterministic Correction}} of {{Qubit Loss}}},
author = {Stricker, R. and Vodola, D. and Erhard, A. and Postler, L. and Meth, M. and Ringbauer, M. and Schindler, P. and Monz, T. and M{\"u}ller, M. and Blatt, R.},
  year = 2020,
  month = sep,
  journal = {Nature},
  volume = {585},
  number = {7824},
  pages = {207--210},
  publisher = {Nature Publishing Group},
  issn = {1476-4687},
  doi = {10.1038/s41586-020-2667-0},
  url = {https://www.nature.com/articles/s41586-020-2667-0},
  urldate = {2024-09-23},
  copyright = {2020 The Author(s), under exclusive licence to Springer Nature Limited}
}

@inproceedings{suchara2015,
  title = {Leakage {{Suppression}} in the {{Toric Code}}},
  booktitle = {2015 {{IEEE International Symposium}} on {{Information Theory}} ({{ISIT}})},
author = {Suchara, M. and Cross, A. W. and Gambetta, J. M.},
  year = 2015,
  month = jun,
  pages = {1119--1123},
  publisher = {IEEE},
  address = {Hong Kong},
  doi = {10.1109/ISIT.2015.7282629},
  url = {http://ieeexplore.ieee.org/document/7282629/},
  urldate = {2025-01-06},
  isbn = {978-1-4673-7704-1}
}

@article{tsai2025,
  title = {Benchmarking and {{Fidelity Response Theory}} of {{High-Fidelity Rydberg Entangling Gates}}},
author = {Tsai, R. B. and Sun, X. and Shaw, A. L. and Finkelstein, R. and Endres, M.},
  year = 2025,
  month = feb,
  journal = {PRX Quantum},
  volume = {6},
  number = {1},
  pages = {010331},
  publisher = {American Physical Society},
  doi = {10.1103/PRXQuantum.6.010331},
  url = {https://link.aps.org/doi/10.1103/PRXQuantum.6.010331},
  urldate = {2025-03-07}
}

@article{wallman2016,
  title = {Noise {{Tailoring}} for {{Scalable Quantum Computation}} via {{Randomized Compiling}}},
author = {Wallman, J. J. and Emerson, J.},
  year = 2016,
  month = nov,
  journal = {Physical Review A},
  volume = {94},
  number = {5},
  pages = {052325},
  publisher = {American Physical Society},
  doi = {10.1103/PhysRevA.94.052325},
  url = {https://link.aps.org/doi/10.1103/PhysRevA.94.052325},
  urldate = {2024-09-23}
}

@article{wu2022,
  title = {Erasure {{Conversion}} for {{Fault-Tolerant Quantum Computing}} in {{Alkaline Earth Rydberg Atom Arrays}}},
author = {Wu, Y. and Kolkowitz, S. and Puri, S. and Thompson, J. D.},
  year = 2022,
  month = aug,
  journal = {Nature Communications},
  volume = {13},
  number = {1},
  pages = {4657},
  publisher = {Nature Publishing Group},
  issn = {2041-1723},
  doi = {10.1038/s41467-022-32094-6},
  url = {https://www.nature.com/articles/s41467-022-32094-6},
  urldate = {2024-09-23},
  copyright = {2022 The Author(s)}
}

@article{xu2024,
  title = {Constant-{{Overhead Fault-Tolerant Quantum Computation}} with {{Reconfigurable Atom Arrays}}},
author = {Xu, Q. and Bonilla Ataides, J. P. and Pattison, C. A. and Raveendran, N. and Bluvstein, D. and Wurtz, J. and Vasi{\'c}, B. and Lukin, M. D. and Jiang, L. and Zhou, H.},
  year = 2024,
  month = jul,
  journal = {Nature Physics},
  volume = {20},
  number = {7},
  pages = {1084--1090},
  publisher = {Nature Publishing Group},
  issn = {1745-2481},
  doi = {10.1038/s41567-024-02479-z},
  url = {https://www.nature.com/articles/s41567-024-02479-z},
  urldate = {2024-09-24},
  copyright = {2024 The Author(s), under exclusive licence to Springer Nature Limited}
}

@misc{yu2024,
  title = {Processing and {{Decoding Rydberg Leakage Error}} with {{MBQC}}},
author = {Yu, C. and Chen, Z. and Deng, Y. and Chen, M. and Lu, C. and Pan, J.},
  year = 2024,
  month = dec,
  eprint = {2411.04664},
  publisher = {arXiv},
  doi = {10.48550/arXiv.2411.04664},
  url = {http://arxiv.org/abs/2411.04664},
  urldate = {2025-01-23},
  archiveprefix = {arXiv}
}

@misc{yu2025,
  title = {Locating {{Rydberg Decay Error}} in {{SWAP-LRU}}},
author = {Yu, C. and Deng, Y. and Chen, M. and Lu, C. and Pan, J.},
  year = 2025,
  month = mar,
  eprint = {2503.01649},
  primaryclass = {quant-ph},
  publisher = {arXiv},
  doi = {10.48550/arXiv.2503.01649},
  url = {http://arxiv.org/abs/2503.01649},
  urldate = {2025-06-18},
  archiveprefix = {arXiv}
}

@misc{ziad2024,
  title = {Local {{Clustering Decoder}}: {{A Fast}} and {{Adaptive Hardware Decoder}} for the {{Surface Code}}},
  shorttitle = {Local {{Clustering Decoder}}},
author = {Ziad, A. B. and Zalawadiya, A. and Topal, C. and Camps, J. and Geh{\'e}r, G. P. and Stafford, M. P. and Turner, M. L.},
  year = 2024,
  month = nov,
  eprint = {2411.10343},
  publisher = {arXiv},
  doi = {10.48550/arXiv.2411.10343},
  url = {http://arxiv.org/abs/2411.10343},
  urldate = {2025-02-06},
  archiveprefix = {arXiv}
}
\bibliographystyle{quantum}
\appendix
\section{Pauli noise channel on the remaining atom}
\label{ap:noise remaining}
In this Appendix, we derive the Pauli noise channel induced on the remaining atom when the other qubit is lost during a CZ gate. 

At the end of the Rydberg pulse, we assume the atom has been reexcited to the Rydberg state $\ket{r}$ and we consider its possible decay back into the computational subspace, described by the decay channel $\mathcal{D}(\rho)$ through the projector
$\Pi = \ket{0}\bra{0} +\ket{1}\bra{1}$ similarly as in ~\cite{jandura2026}:
\begin{equation}
    \mathcal{D}(\rho) = \Pi\rho\Pi + \bra{r}\rho\ket{r}\Pi/2.
\end{equation}
Applied to our case, this becomes
\begin{eqnarray}
    \mathcal{D}(\rho) &=& \ket{0}\bra{0}\rho\ket{0}\bra{0} + \frac{\ket{0}\bra{r}\rho\ket{r}\bra{0}}{2} \nonumber\\
    &+& \frac{\ket{1}\bra{r}\rho\ket{r}\bra{1}}{2}.
\end{eqnarray}
By subsequently identifying the Rydberg state with the original state $\ket{1}$ , we can apply the decay map onto the original density matrix state $\tilde\rho$ prior to the CZ pulse that allows us to express the Kraus operators in the Pauli basis:
\begin{align}
        \mathcal{D}(\tilde{\rho})   &= \frac{\mathbb{I}+Z}{2}\tilde{\rho}\frac{\mathbb{I}+Z}{2} +\frac{1}{2} \frac{X+iY}{2}\tilde{\rho}\frac{X-iY}{2} \nonumber\\
        &+ \frac{1}{2}\frac{\mathbb{I}-Z}{2}\tilde{\rho}\frac{\mathbb{I}-Z}{2}
\end{align}
Applying the Pauli twirling approximation (PTA) which consists in keeping only the diagonal elements of the noise channel we obtain:
\begin{align}
        \mathcal{D}(\tilde{\rho}) &= \frac{3}{8}\left(\tilde{\rho} + Z\tilde{\rho}Z\right)+\frac{1}{8}\left(X\tilde{\rho}X + Y\tilde{\rho}Y\right)
\end{align}

Note that PTA is an approximation but can be made exact for Clifford two-qubit gates by randomly dressing them with Pauli gates~\cite{kern2005,wallman2016,hashim2021}. 

\section{State-dependency of loss channels}
\label{ap:state dependent}
In order to be accurate, loss channels should be implemented taking into account that it is much more likely to lose an atom when the latter is in state $\ket{1}$ because of Rydberg excitations of that state during the pulse of the CZ gate, as shown in Sec.~\ref{sec:noise model}. Our current implementation of the loss channel is state independent. In this Appendix we analyze the impact of such approximation in the context of error correction.

For the special case of quantum error correction codes where QEC cycles can only be implemented via Clifford operation such as CSS codes, the logical states $\ket{0}_L$ and $\ket{1}_L$ are an even superposition of physical states $\ket{0}$ and $\ket{1}$:
\begin{equation}
\ket{0/1}_L =\frac{\ket{0}\ket{\psi_0} +\ket{1}\ket{\psi_1}}{\sqrt{2}}
\label{eq:clifford state}
\end{equation}
where in the right-hand side of the equation, the first ket can represent any qubits in the system (say $i$) and $\ket{\psi_0}$ (resp. $\ket{\psi_1}$) represents the normalized wavefunction on the rest of the qubits if the qubit $i$ is in $\ket{0}$ (resp. in $\ket{1}$). The underlying reason being that Clifford states are either similar to Eq.~\eqref{eq:clifford state} or pure $\ket{0}$ or $\ket{1}$ in which case the latter would imply that the qubit is disentangled from the rest of the system. However, in that case, the disentangled qubit (e.g. ancilla qubits at the beginning of each QEC cycle or LDU), are always initialized in the $\ket{+/-}$ prior to the CZ gate application in order for the latter to act non-trivially on it. As a consequence, any random superposition of $\ket{0}_L$ and $\ket{1}_L$ is evenly balanced between $\ket{0}$ and $\ket{1}$ for any qubit. We go further and postulate that for any pair of neighbouring data and ancilla qubits the state before applying a CZ gate can be written in the following form:
\begin{equation}
\frac{\ket{00}\ket{\psi_{00}} + \ket{01}\ket{\psi_{01}} + \ket{10}\ket{\psi_{10}} + \ket{11}\ket{\psi_{11}}}{2}.
\end{equation}

In addition, introducing Pauli noise channel does not affect this balance. In that case,  the loss channels act uniformly over any qubits. If we assume losses are negligible when the qubit is in $\ket{0}$ and that the loss probability $p_l$ of the CZ gates has been computed when both qubits were in state $\ket{11}$, then the latter should be uniformly rescaled by a factor $3/4$ . 

However, as we explain below, the loss of one qubit may modify the loss probability of other qubits in a state-dependent manner that is not captured by our simulation.

Consider two neighbouring boundary data qubits $i$ and  $j$ stabilized by $Z_iZ_j$. Before the error-correction cycle begins, the joint state of the system can be written as
\begin{equation}
    \ket{\psi'}=\frac{\ket{00}_{i,j}\ket{\psi_{00}} + \ket{11}_{i,j}\ket{\psi_{11}}}{\sqrt{2}}.
\end{equation}

If data qubit $i$ is lost during its first interaction with an ancilla, the system is projected onto
\begin{equation}
    \ket{\psi'}=\ket{l1}_{i,j}\ket{\psi_{11}}
\end{equation} where $\ket{l}$ denotes the lost qubit.
As a consequence, during the subsequent CZ gate between the ancilla and the second data qubit $j$, the probability of losing $j$ is effectively increased, because the state of $j$ has been projected onto $\ket{1}$.

Our simulation does not account for such state-dependent loss effect. We leave it for future investigation as well as the potential modification of the decoding strategy.
\end{document}